\documentclass[english,traditabstract]{aa}

\usepackage{amsmath}
\usepackage{amssymb}
\usepackage{txfonts}
\usepackage{epsfig,graphicx}
\usepackage[svgnames]{xcolor}
\usepackage[colorlinks=true, linkcolor=Blue, citecolor=Blue, urlcolor=Blue]{hyperref}
\usepackage{natbib}


\bibpunct{(}{)}{;}{a}{}{,} 
\usepackage{graphicx}
\usepackage{amsmath}
\defcitealias{2012arXiv1202.5242R}{RM12}
\defcitealias{2006MNRAS.365..414B}{B06}

\makeatother

\begin{document}

\title{Analytical shear and flexion of Einasto dark matter haloes}

\author{E. Retana-Montenegro \inst{1} 
\and F. Frutos-Alfaro \inst{1}
\and M. Baes \inst{2}}

\institute{Escuela de F\'{i}sica, Universidad de Costa Rica, San Pedro 11501, Costa Rica\\
\email{edwin@fisica.ucr.ac.cr}
\and Sterrenkundig Observatorium, Universiteit Gent, Krijgslaan 281-S9, B-9000 Gent, Belgium\\
\email{maarten.baes@ugent.be}
}

\date{Received ...; Accepted...}

\abstract{\noindent $N$-body simulations predict that dark matter haloes are
described by specific density profiles on both galactic- and cluster-sized
scales. Weak gravitational lensing through the measurements of their
first and second order properties, shear and flexion, is a powerful
observational tool for investigating the true shape of these profiles.
One of the three-parameter density profiles recently favoured in the
description of dark matter haloes is the Einasto profile. We present
exact expressions for the shear and the first and second flexions of Einasto
dark matter haloes derived using a Mellin-transform formalism in terms
of the Fox $H$ and Meijer $G$ functions, that are valid for general
values of the Einasto index. The resulting expressions can be written
as series expansions that permit us to investigate the asymptotic
behaviour of these quantities. Moreover, we compare the shear and
flexion of the Einasto profile with those of different mass profiles
including the singular isothermal sphere, the Navarro-Frenk-White
profile, and the S\'ersic profile. We investigate the concentration
and index dependences of the Einasto profile, finding that the shear
and second flexion could be used to determine the halo concentration,
whilst  for the Einasto index the shear and first and second flexions
may be employed. We also provide simplified expressions for the weak
lensing properties and other lensing quantities in terms of the generalized
hypergeometric function.}


\maketitle

\section{Introduction}

\label{sec:Intro}

A more accurate description of the elements that constitute our universe,
such as the dark matter haloes that are believed to exist around galaxies
and clusters, is of crucial importance for our understanding of cosmological
structural formation. Recent results from $N$-body simulations of
cold dark matter (CDM) \citep{2004MNRAS.349.1039N,2006AJ....132.2685M,2008MNRAS.387..536G,2008MNRAS.388....2H,2009MNRAS.398L..21S,2010MNRAS.402...21N,2011MNRAS.415.3177R,2012arXiv1202.6061V}
indicate that nonsingular three-parameter models such as the \citet{1965TIAAA.17..01}
profile, fit a wide range of dark matter haloes better than singular
two-parameter models, e.g. the Navarro, Frenk, and White (NFW) profile
\citep*{1996ApJ...462..563N,1997ApJ...490..493N}.

\noindent The Einasto profile is given by

\noindent 
\begin{equation}
\rho(r)=\rho_{\text{s}}\exp\left\{ -d_{n}\left[\left(\frac{r}{{r_{{\text{s}}}}}\right)^{1/n}-1\right]\,\right\} ,\label{eq:einasto_generic}
\end{equation}

\noindent where $r$ is the spatial radius, the shape parameter $n$
is called the Einasto index, ${r_{{\text{s}}}}$ represents the radius
of the sphere that contains half of the total mass, $\rho_{\text{s}}$
is the mass density at $r={r_{{\text{s}}}}$, and $d_{n}$ is a function
that ensures that ${r_{{\text{s}}}}$ is indeed the half-mass radius.
An analytical expansion for the function $d_{n}\approx3n-1/3+8/1215n+\mathcal{O}\left(n^{2}\right)$
is provided by Retana-Montenegro et al.  \citep[][hereafter \citetalias{2012arXiv1202.5242R}]{2012arXiv1202.5242R}.
One important characteristic of this profile is that its power-law
logarithmic slope, $\gamma\left(r\right)=-{\text{d}}\text{ln}\rho/{\text{d}}\text{ln}r\sim r^{1/n}$,
depends on the Einasto index, which provides a profile that
more accurately fits in the inner regions of simulated dark matter
haloes than other profiles such as the NFW profile. In the study
of real galaxies, several authors have used multi-component Einasto
models, consisting generally of two or more Einasto components for
each galaxy, where each component represents a homogeneous stellar
population with its own set of parameters. For example, some of the
first galaxies to be modelled using multi-component Einasto models
were M31 by \citet{1969Afz.....5..137E} with values of $0.25\leq n\leq1$,
and other nearby galaxies such as Milky Way, M87, M32, Fornax, and Sculptor,
and M31 by \citet{1974smws.conf..291E} with $0.5\leq n\leq4$. Later,
in a series of papers multi-component Einasto models were employed
to model the luminous components of several galaxies such as the Milky
Way \citep{1989A&A...223...89E}, M87 \citep{1991A&A...248..395T},
M31 \citep{1994A&A...286..753T}, and M81 \citep{1998A&A...335..449T};
in these papers, the Einasto index is characterised by values of $0.36\leq n\leq7.1$.
The seven distant spiral galaxies GSS 074-2237, GSS 064-4412, GSS
094-2210, GSS 104-4024, GSS 064-4442, MDS uem0-043, and HDFS J223247.66-603335.9
were studied by \citet{2003A&A...403..529T} and \citet{2005A&A...433...31T},
respectively. As in the earlier works mentioned, they modelled each
visual component with a Einasto profile, the authors found values
of $0.25\leq n\leq0.91$ and noted that the Einasto indices for the
disk component of the galaxies at high redshift follow a trend of
having smaller values than the ones at lower redshift. \citet{2006MNRAS.371.1269T}
fitted a multi-component Einasto model to the Sombrero galaxy, with
$0.78\leq n\leq3$ for the visual components. \citet{2007arXiv0707.4375T}
and \citet{2007arXiv0707.4374T} presented a multi-component Einasto
law study of M31: using photometric data and metallicity measurements,
they obtained the matter distribution of luminous components with
$0.70\leq n\leq4.20$, then tried to fit several models for the dark
matter halo using kinematical data from the literature
to construct a dynamical model and derive the dark matter density
of the galaxy, they concluded that Einasto and NFW profiles give the
best fits. \citet{2011arXiv1112.3120D} fitted the surface brightness
density of a sample of elliptical galaxies using a multi-component
Einasto profile, finding values of $1\leq n\leq3$ for the central
components, and $5\leq n\leq8$ for the outer components. \citet{2011AJ....142..109C},
who studied the rotation curves of low mass spiral galaxies, modelled
the dark matter halo with a Einasto profile, and obtained smaller
values of $n$ than predicted by computational simulations. On the
other hand, according to $N$-body numerical calculations the Einasto
index dependens on both the halo mass and redshift \citep{2008MNRAS.388....2H,2008MNRAS.387..536G}.
Typical values of the Einasto index are in the range $5\leq n\leq8$
according to the results of $N$-body simulations \citep{2004MNRAS.349.1039N,2008MNRAS.387..536G,2008MNRAS.388....2H,2010MNRAS.402...21N}.
\citet{2012arXiv1202.6061V} analysed dark matter haloes of Milky
Way-like systems and concluded that the Einasto model with values
of $2\leq n\leq5$ is preferred over the NFW profile.

An alternative form of the density often used in dark matter halo
studies is

\noindent 
\begin{equation}
\rho\left(r\right)=\rho_{-2}\,\exp\left\{ -2n\left[\left(\frac{r}{r_{-2}}\right)^{1/n}-1\right]\,\right\} ,\label{eq:einasto_halo_vers}
\end{equation}

\noindent where $r_{-2}$ is the radius at which the logarithmic slope
of the density distribution has a value of $-2$ and $\rho_{-2}=\rho\left(r_{-2}\right)$.
A useful quantity to define is the concentration $c_{E}=r_{200}/r_{-2}$,
where $r_{200}$ is the virial radius of a halo of mass $M_{200}$,
whose density is 200 times the critical density of the Universe at
the halo redshift. One of the advantages of the Einasto profile over
other profiles is that it has excellent agreement with the conditions
outlined by \citet{1969AN....291...97E} for constructing real galactic
models,\textbf{ }specifically, some moments must be finite. In particular,
for this profile some moments, such as the total mass, central gravitational
potential, and effective radius, are finite. In contrast, other profiles
have logarithmic moments that must be truncated at some radius to ensure that the profile remains finite.

Gravitational lensing provides a direct way to study the mass distribution
of large structures in the universe, such as galaxies and clusters,
without making any assumptions about their dynamical state or composition.
Lensing studies taking advantage of high\textendash{}quality imaging
have proven to be successful in mapping the distribution of dark matter
in clusters and galaxies \citep{2003ApJ...598..804K,2006ApJ...648L.109C,2008ApJ...687..959B,2009ApJ...702..603A,2010PASJ...62..811O,2010MNRAS.405.2215O,2011ApJ...741..116O,2011A&A...529A..93H,2012ApJ...744...94R,2012ApJ...747...96J,MNR:MNR20248}.
There are two lensing regimes: the strong regime, where multiple images
or strong distortions of a galaxy can be produced by an intervening
distribution of matter, and the weak regime, where the lensed galaxy
image is only slightly distorted, causing the intrinsic elliptical
galaxy to appear as a distorted elliptical image. Weak lensing is a
valuable and accurate tool for determining the shapes of dark-matter-halo
density profiles, such as ellipticity \citep{2004ApJ...606...67H,2006MNRAS.370.1008M,2007ApJ...669...21P,2009ApJ...695.1446E,2010ApJ...721..124D,2010MNRAS.405.2215O}
and triaxiality \citep{2005ApJ...632..841O,2005A&A...443..793G,2007MNRAS.380..149C,2009MNRAS.393.1235C,2012MNRAS.420..596F}.
Until now, most weak lensing studies have considered only linear-order
effects such as the weak shear, the quantity responsible for the
induced ellipticity in the galaxy (see e.g. \citet{1995ApJ...449..460K,2001PhR...340..291B,9783540303091,2008ARNPS..58...99H}
for reviews). 

In the past few years, the study of high-order lensing properties
has grown in importance (\citealp{2002ApJ...564...65G}, \citealp{2005ApJ...619..741G},
\citet[][hereafter \citetalias{2006MNRAS.365..414B}]{2006MNRAS.365..414B}).
These properties written as high-order derivatives of the deflection
potential can be recognized as convergence and shear gradients. The
convergence gradient, called the first flexion $\mathcal{F}$, induces
a centroid shift in the lensed image with respect to the source or
{}``skewness''. The shear gradient, called second flexion $\mathcal{G}$,
generates an arc-like shape in the lensed image or {}``arcness''.
Weak flexion provides useful information about dark matter haloes
on galactic- and cluster-sized scales, particularly when probing substructure
on smaller-scales where flexion is more sensitive to shear-only studies
\citep{2009MNRAS.395.1438L,2010MNRAS.409..389B,2010arXiv1008.3088E}. 

Several methods have been developed to measure the flexion of a lensed
image, e.g. shapelets (\citetalias{2006MNRAS.365..414B}; \citealt{2007MNRAS.380..229M};
\citealt{2012MNRAS.tmp.2376F}) and surface brightness moments \citep{2006ApJ...645...17I,Irwin200583,Irwin2007,2007ApJ...660.1003G,2007ApJ...660..995O,2008ApJ...680....1O,2009ApJ...699..143O,2008A&A...485..363S}.
\citet{2011ApJ...736...43C} introduced a new method, called the analytic
image model (AIM), to study flexion in astronomical images. Observational
measurements of flexion include the detection of mass substructure
in the Abell 1689 cluster by \citet{2007ApJ...666...51L} and \citet{2011ApJ...736...43C}
using observations of the Hubble Space Telescope (HST), as well as
\citet{2008ApJ...680....1O} employing Subaru images; galaxy-galaxy
flexion detection in the ground-based survey Deep Lens Survey \citep{2005ApJ...619..741G}
and the space-based HST COSMOS survey \citep{2011MNRAS.412.2665V}.

In addition, flexion has been proposed as a powerful cosmological
tool: \citet{MNR:MNR17838} suggested the use of convergence shear
and flexion maps to decrease errors in the measuring standard candles
distances, \citet{2011arXiv1104.3955C} studied how the flexion signal-to-noise
ratio could be used to discern between cosmological models, \citet{MNR:MNR17838}
and \citet{MNR:MNR20051} proposed the use of cosmic flexion to probe
large-scale structure. \citet{2009MNRAS.400.1132H} studied the halo
ellipticity on galactic scales, and found that the inclusion of flexion
yields tighter constraints on ellipticity than shear-only studies.
\citet{2011A&A...528A..52E} and \citet{2011MNRAS.417.2197E} proposed
a new way to determine the halo ellipticity using the ratio of tangential-to-radial
flexion and studied its behaviour as a radius function. \citet{2012MNRAS.421.1443E}
concluded that flexion is more sensitive to ellipticity than shear
by performing a likehood analysis of mock flexion and shear data.
Additionally, \citet{2012MNRAS.419.2215V} considered the case in
which cross-terms between both shear and flexion and between intrinsic
galaxy ellipticities and flexion are not ignored, concluding that
these terms can cause a considerable bias in the flexion estimations. 

In view of the increased use of the Einasto profile in cosmological
studies (e.g. see \citet{2010JCAP...08..004C,2011MNRAS.415.3177R,2011AJ....142..109C,2011arXiv1112.3120D,2011arXiv1111.3556C,2012JCAP...05..016N}),
it is natural to extend its applications to weak shear and flexion
lensing studies. Previously, several authors had performed weak lensing
studies using the Einasto profile. For example, \citet{2008MNRAS.388....2H}
measured the cross-correlations between halo centres and mass, and
between galaxies and mass, in the Millennium Run \citep{2005Natur.435..629S},
and found that the Einasto profile provides a close fit in the inner
regions of their two-part model of the halo-mass cross-correlation
function. \citet{1475-7516-2008-08-006} analysed, using a weak statistical
approach, a sample of galactic- and cluster-sized dark matter haloes
from the Sloan Digital Sky Survey, and obtained very similar concentration-mass
relations for the NFW and Einasto profiles. \citet{2010A&A...520A..30M}
used analytical approximations of the shear of the Einasto profile
to compare it with the NFW shear.

Parametric models such as the singular isothermal sphere, the NFW,
and S\'ersic profiles have been used to model the dark matter distribution
in weak lensing analyses (e.g. \citealt{2009MNRAS.400.1132H,2011ApJ...729..127U,2011A&A...534A..14V,2011MNRAS.417.2197E,2012MNRAS.419.2215V,2012A&A...540A..61S}),
the properties of these models having been studied by several authors
(\citealt{2000ApJ...534...34W}; \citetalias{2006MNRAS.365..414B};
\citealt{2009MNRAS.396.2257L}). In the case of the Einasto profile,
\citetalias{2012arXiv1202.5242R} studied the analytical properties
of the Einasto profile by applying a Mellin-transform formalism. In
terms of Fox $H$ and Meijer $G$ functions, they derived analytical
expressions of lensing properties for all\emph{ }values of the Einasto
index, concentrating on the surface mass density, cumulative mass,
deflection angle, and deflection potential. However, by means of the
Mellin-transform formalism it is possible to extensively study the
weak-lensing analytical properties of the Einasto profile. This study
provides analytical expressions that can used to model realistic Einasto
dark matter haloes in weak lensing modelling studies.

In this work, we apply Mellin-transform formalism to obtain and study
in detail the analytical expressions for the weak lensing properties
of the Einasto profile: the shear, and first, and second flexions. This
paper is organized as follows. We summarize the weak lensing formalism
in Section \ref{sec:Section2-1}, and present the Mellin-transform
technique in Section \ref{sec:Section2-2}. In Section \ref{sec:Section3}
we derive closed expressions for the shear and first and second flexions
in terms of the Fox $H$ and Meijer $G$ functions. We then use the
series expansions of these expressions to investigate their asymptotic
behaviour. In Section \ref{sec:Section4}, we compare our results
with those for the SIS, NFW, and S\'ersic profiles. In Section \ref{sec:Section5},
we summarise and discuss our main results. Finally, in the appendices
\ref{Appendix-A} and \ref{Appendix-B} we provide series expansions
of the lensing properties, and explicit expressions in terms of the
generalized hypergeometric function, respectively. Throughout the
paper, we adopt a cosmological model with the matter density $\Omega_{M}=0.26$,
the cosmological constant $\Omega_{\Lambda}=0.74$, and the Hubble
constant $H_0=72\, {\rm km}\,{\rm s}^{-1}{\rm Mpc}^{-1}$.

\section{Theory}

We provide a brief description of the two main theoretical aspects
employed throughout this paper.

\subsection{Weak lensing formalism \label{sec:Section2-1}}

\noindent The weak lensing formalism using complex notation was introduced
by \citetalias{2006MNRAS.365..414B}. In the thin lens approximation,
the lens equation is given by \citep{1992grle.book.....S} 

\begin{equation}
\boldsymbol{\beta}=\boldsymbol{\theta}-\boldsymbol{\nabla}\psi(\boldsymbol{\theta}),\label{eq:lens-equation}
\end{equation}

\noindent where $\boldsymbol{\beta}$ and $\boldsymbol{\theta}$ denote
the positions on the source plane, and on the image plane, respectively,
and $\psi(\boldsymbol{\theta})$ is the deflection potential defined
by a two-dimensional Poisson, $\boldsymbol{\nabla^{2}}\psi(\boldsymbol{\theta})=2\kappa(\boldsymbol{\theta})$,
with the convergence $\kappa(\boldsymbol{\theta})$. Moreover, the
convergence can be written as 

\begin{equation}
\kappa(\boldsymbol{\theta})=\frac{\Sigma(\boldsymbol{\theta})}{\Sigma_{{\text{crit}}}},\label{eq:convergence}
\end{equation}

\noindent where $\Sigma(\boldsymbol{\theta})$ is the surface mass
density, 

\begin{equation}
\Sigma_{{\text{crit}}}=\frac{c^{2}\, D_{\text{S}}}{4\pi\, G\, D_{\text{L}}\, D_{\text{LS}}}
\end{equation}

\noindent is the critical surface mass density, and $D_{\text{L}}$,
$D_{\text{S}}$, and $D_{\text{LS}}$ are the angular distances from
observer to lens, from observer to source, and from lens to source,
respectively. In addition, it is convenient to define the complex
gradient operator (\citetalias{2006MNRAS.365..414B})

\noindent 
\begin{equation}
\partial=\frac{\partial}{\partial\theta_{1}}+i\,\frac{\partial}{\partial\theta_{2}}=\partial e^{i\phi},
\end{equation}
where $\phi$ is the rotation angle, relative to the basis. The $\partial$
operator is simply a spin-$s$%
\footnote{We define a spin-$s$ lensing quantity by requiring that it is invariant
under rotations $\phi=2\pi/s$, where $s$ is any natural number except
zero.%
} raising operator and its complex conjugate $\partial^{\star}$ a
spin-$s$ lowering operator.

\noindent When we study gravitational lensing on scales where the
deflection potential changes are larger than the scale of the lensed
image, we can expand up to second order the lens equation in eq. (\ref{eq:lens-equation})
around the neighbourhood of the lensed image%
\footnote{We do not consider crossed terms in the lens equation expansion. %
} 

\begin{equation}
\beta_{i}=\mathcal{A}_{ij}\theta_{j}+\frac{1}{2}\mathcal{D}_{ijk}\theta_{j}\theta_{k},
\end{equation}

\noindent where $\mathcal{A}_{ij}$ is the Jacobian matrix defined
by

\begin{equation}
\mathcal{A}_{ij}=\frac{\partial\beta_{i}}{\partial\theta_{j}}=\left(\begin{array}{cc}
1-\kappa-\gamma_{1} & -\gamma_{2}\\
-\gamma_{2} & 1-\kappa+\gamma_{1}
\end{array}\right),
\end{equation}

\noindent with the convergence, 
\begin{equation}
\kappa=\frac{1}{2}\left(\psi_{xx}+\psi_{yy}\right)=\frac{1}{2}\partial\partial^{\star}\psi,\label{eq:convergence-pot}
\end{equation}

\noindent which is a spin-0 field, and with $\gamma_{1}=\frac{1}{2}\left(\psi_{xx}+\psi_{yy}\right)$,$\gamma{}_{2}=\psi_{xy}$,
the components of the complex shear

\noindent 
\begin{equation}
\gamma=\gamma_{1}+i\,\gamma_{2}=\left|\gamma\right|e^{2i\phi}=\frac{1}{2}\partial\partial\psi,\label{eq:shear-pot}
\end{equation}

\noindent which is a spin-2 field. The matrix $\mathcal{D}_{ijk}=\partial A_{ij}/\partial\theta_{k}$
describes the behaviour of the convergence and shear across the lensed
image by introducing two new lensing properties

\begin{equation}
\mathcal{D}_{ijk}=\mathcal{F}_{ijk}+\mathcal{G}_{ijk},
\end{equation}

\noindent namely the first flexion field or spin-1 first flexion $\mathcal{F}_{ijk}$
and the second flexion field or spin-3 second flexion $\mathcal{G}_{ijk}$.
Both field components can be expressed as third-order derivatives
of the deflexion potential \citep{2009MNRAS.400.1132H}

\begin{equation}
\mathcal{F}_{1}=\frac{1}{2}\left(\psi_{xxx}+\psi_{yyx}\right),
\end{equation}
\begin{equation}
\mathcal{F}_{2}=\frac{1}{2}\left(\psi_{xxy}+\psi_{yyy}\right),
\end{equation}

\begin{equation}
\mathcal{G}_{2}=\frac{1}{2}\left(\psi_{xxx}-3\psi_{xyy}\right),
\end{equation}

\begin{equation}
\mathcal{G}_{2}=\frac{1}{2}\left(3\psi_{xxy}-\psi_{yyy}\right),
\end{equation}

\noindent and taking advantage of the complex formalism, we can compactly
write the first and second flexions as

\noindent 
\begin{equation}
\mathcal{F}=\mathcal{F}_{1}+i\,\mathcal{F}_{2}=\frac{1}{2}\partial\partial\partial^{\star}\psi,\label{eq:1st-flex-pot}
\end{equation}

\noindent 
\begin{equation}
\mathcal{G}=\mathcal{G}_{1}+i\,\mathcal{G}_{2}=\frac{1}{2}\partial\partial\partial\psi.\label{eq:2nd-flex-pot}
\end{equation}

\noindent From eqs. (\ref{eq:1st-flex-pot}) and (\ref{eq:2nd-flex-pot}),
one can clearly see the rotation symmetry for both flexions. Applying
the complex conjugate operator to eqs. (\ref{eq:convergence-pot}),
and (\ref{eq:shear-pot}) and comparing with eqs. (\ref{eq:1st-flex-pot}),
and (\ref{eq:2nd-flex-pot}), we find a compact and elegant definition
of the second-order properties as gradients of the first-order lensing
properties

\noindent 
\begin{equation}
\mathcal{F}=\left|\mathcal{F}\right|e^{i\phi}=\partial\kappa=\partial^{\star}\gamma,\label{eq:1st-flex-pot-1}
\end{equation}

\noindent 
\begin{equation}
\mathcal{G}=\left|\mathcal{\mathcal{G}}\right|e^{3i\phi}=\partial\gamma.\label{eq:2nd-flex-pot-1}
\end{equation}

\subsection{Mellin-transform technique \label{sec:Section2-2}}

The Mellin transform technique \citep{marichev1983handbook,1996MER.16..05,159829184X}
consists in that one-dimensional definite integrals 

\begin{equation}
f(z)=\int_{0}^{\infty}g(t,z)\,{\text{d}}t,\label{Marichev}
\end{equation}

\noindent can be expressed as the Mellin convolution of the functions
$f_{1}$ and $f_{2}$

\begin{equation}
f(z)=\int_{0}^{\infty}f_{1}(t)\, f_{2}\left(\frac{z}{t}\right)\,\frac{{\text{d}}t}{t}.\label{eq:mellin-conv}
\end{equation}

\noindent The Mellin convolution theorem, which states that the Mellin
transform of a Mellin convolution of two functions is the pointwise
product of their Mellin transforms, can be applied to eq. (\ref{eq:mellin-conv})
inverting the Mellin transform of the Mellin convolution, $f\left(z\right)$
can be expressed as the inverse Mellin transform of the pointwise
product of the $f_{1}$ and $f_{2}$ Mellin transforms. The Mellin
transform is defined by
\begin{equation}
{\mathfrak{M}}_{f}(u)=\phi(u)=\int_{0}^{\infty}f(z)\, z^{u-1}\,{\text{d}}z,
\end{equation}

\noindent and the inverse Mellin transform by

\begin{equation}
{\mathfrak{M}}_{\phi}^{-1}(z)=f(z)=\frac{1}{2\pi i}\int_{\mathcal{L}}\phi(u)\, z^{-u}\,{\text{d}}u,
\end{equation}

\noindent where the integration path is a vertical line in the complex
plane. 

\noindent The integral in eq. (\ref{Marichev}) may then be written
as 

\begin{equation}
f(z)=\frac{1}{2\pi i}\int_{\mathcal{L}}{\mathfrak{M}}_{f_{1}}(u)\,{\mathfrak{M}}_{f_{2}}(u)\, z^{-u}\,{\text{d}}u.\label{MellinBarnes}
\end{equation}

\noindent With the requirement that $f_{1}$ and $f_{2}$ are of hypergeometric
type, their Mellin transforms can be written as products of the form
$\Gamma\left(a+Au\right)$ or $\left[\Gamma\left(a+Au\right)\right]^{-1}$,
with $\Gamma\left(v\right)$ the gamma function and $A$ real. The
resulting integral in eq. (\ref{MellinBarnes}) is of the Mellin-Barnes
type and it then can be evaluated as either a Fox $H$ function for
$A\neq1$ or as a Meijer $G$ function for $A=1$.

\section{Weak lensing \label{sec:Section3}}

We derive closed expressions for the weak-lensing first- and second-order
properties of the Einasto profile: the shear $\gamma$, and the first
$\mathcal{F}$ and second $\mathcal{G}$ flexions in terms of Fox
$H$, Meijer $G$ functions, and the generalized hypergeometric function.
Using these expressions, we calculate the expansion series and investigate
its asymptotic behaviour. The results of this section provide a useful
and straightforward way to study weak lensing, where the matter distribution
is believed to be described by an Einasto profile.

\subsection{Convergence and shear \label{sec:Section3-1}}

In the weak lensing regime up to first order, the lensed galaxy image
has two distortions: the convergence $\kappa$ causes an isotropic
stretching in the lensed image, which magnifies the image by increasing
its size, and the shear $\gamma$ also causes an anisotropic stretching
in the lensed image, that is responsible for the induced ellipticity
in the lensed galaxy. 

\noindent Foremost, to calculate the convergence, we must project
the density profile on the lens plane using an Abel transform

\begin{equation}
\Sigma\left(\xi\right)=2\,\int_{\xi}^{\infty}\frac{\rho\left(r\right)\, r\,{\text{d}}r}{\sqrt{r^{2}-\xi^{2}}},\label{eq:sigma}
\end{equation}

\noindent where $\xi$ is the radius from the lens centre, and $r$
is the spatial radius. We follow here the notation of \citetalias{2012arXiv1202.5242R}
for the Einasto profile

\noindent 
\begin{equation}
\rho(r)=\rho_{0}\exp\left[-\left(\frac{r}{h}\right)^{1/n}\right],\label{eq:einasto_gen_vers}
\end{equation}

\noindent where we define the central density $\rho_{0}=\rho_{s}\, e^{d_{n}}=\rho_{-2}\, e^{2n}$
and scale length $h={r_{{\text{s}}}}/d_{n}^{n}=r_{-2}/\left(2n\right)^{n}$.
Additionally, we define another quantity, the central convergence

\begin{equation}
\kappa_{\text{c}}\equiv\frac{\Sigma\left(0\right)}{\Sigma_{{\text{crit}}}}=\frac{2\,\rho_{0}\, h\, n\,\Gamma\left(n\right)}{\Sigma_{{\text{crit}}}}.\label{eq:central-convergence}
\end{equation}

\noindent Combining eqs. (\ref{eq:convergence}), (\ref{eq:sigma}),
(\ref{eq:einasto_gen_vers}) and (\ref{eq:central-convergence}),
we get 

\begin{equation}
\kappa\left(x\right)=\frac{\kappa_{\text{c}}}{n\,\Gamma\left(n\right)}\,\int_{x}^{\infty}\frac{{\text{e}}^{-s^{1/n}}\, s\,{\text{d}}s}{\sqrt{s^{2}-x^{2}}},\label{eq:Convergence_integral}
\end{equation}

\noindent where $x=\theta\, D_{\text{L}}/h=\xi/h$ and $s=r/h$ are
the dimensionless radii. 

\noindent The integral in eq. (\ref{eq:Convergence_integral}) cannot
be expressed in terms of elementary or special functions for general
values of $n$. However, using the Mellin transform technique explained
in Section (\ref{sec:Section2-2}), we can write this integral as
a Mellin-Barnes integral 

\begin{equation}
\kappa\left(x\right)=\frac{\kappa_{\text{c}}\,\sqrt{\pi}\, x\,}{\Gamma\left(n\right)}\,\frac{1}{2\pi i}\int_{\mathcal{L}}\frac{\Gamma\left(2ny\right)\Gamma\left(-\frac{1}{2}+y\right)}{\Gamma\left(y\right)}\left[\, x^{2}\right]^{-y}{\text{d}}y.\label{eq:Convergence-MellinBarnes}
\end{equation}

\noindent The Fox $H$ function \citep{1961TAMS...98..395} is denoted
as a Mellin-Barnes integral, 
\begin{multline}
H_{p,q}^{m,n}\left[\left.\begin{matrix}({\boldsymbol{a}},{\boldsymbol{A}})\\
({\boldsymbol{b}},{\boldsymbol{B}})
\end{matrix}\,\right|\, z\right]=\\
\frac{1}{2\pi i}\int_{{\cal L}}\frac{\prod_{j=1}^{m}\Gamma(b_{j}+B_{j}s)\prod_{j=1}^{n}\Gamma(1-a_{j}-A_{j}s)}{\prod_{j=m+1}^{q}\Gamma(1-b_{j}-B_{j}s)\prod_{j=n+1}^{p}\Gamma(a_{j}+A_{j}s)}\, z^{-s}\,{\text{d}}s.\label{eq:defH}
\end{multline}
Comparing the integral in eq. (\ref{eq:Convergence-MellinBarnes})
with the above definition, we obtain a close expression for the convergence
in terms of the Fox $H$ function \citepalias{2012arXiv1202.5242R} 

\noindent \begin{flushleft}
\begin{equation}
\kappa\left(x\right)=\frac{\kappa_{\text{c}}\,\sqrt{\pi}}{\Gamma\left(n\right)}\, x\, H_{1,2}^{2,0}\left[\begin{array}{c}
(0,\,1)\\
(0,\,2n),(-\frac{1}{2},\,1)
\end{array}\biggr|\, x^{2}\right].\label{eq:convergence-Fox}
\end{equation}

\par\end{flushleft}

\noindent The shear for an circularly symmetric lens is \citep{1991ApJ...370....1M}
\begin{equation}
\gamma\left(x\right)=\bar{\kappa}\left(x\right)-\kappa\left(x\right),\label{eq:shear}
\end{equation}

\noindent where 

\noindent 
\begin{equation}
\bar{\kappa}(x)=\frac{2}{x^{2}}\int_{0}^{x}x'\,\kappa(x')\,{\text{d}}x',\label{eq:mean-kappa}
\end{equation}

\noindent is the average convergence within the dimensionless radius
$x$.

\noindent Inserting eq. (\ref{eq:convergence-Fox}) into eq. (\ref{eq:mean-kappa})
and substituting this result along with eq. (\ref{eq:convergence-Fox})
into eq. (\ref{eq:shear}), we can re-express the resulting integral
as a Fox $H$ function and obtain the shear for the Einasto profile

\noindent \begin{flushleft}
\begin{equation}
\gamma\left(x\right)=\left\{ \frac{\kappa_{\text{c}}\,\sqrt{\pi}}{\Gamma\left(n\right)}\, x\, H_{2,3}^{2,1}\left[\begin{array}{c}
(-\frac{1}{2},\,1),(0,\,1)\\
(0,\,2n),(\frac{1}{2},\,1),(-\frac{3}{2},\,1)
\end{array}\biggr|\, x^{2}\right]\right\} e^{2i\phi}.\label{eq:shear-Fox}
\end{equation}

\par\end{flushleft}

\noindent Eq. (\ref{eq:shear-Fox}) provides an expression for the
shear in terms of one Fox $H$ function, instead of two Fox $H$ functions
as found by \citetalias{2012arXiv1202.5242R}. Writing eq. (\ref{eq:shear-Fox})
in terms of one instead of two Fox $H$ functions makes it easier
to manipulate for analytical and numerical purposes.

\noindent The Fox $H$ function is a very general function and reduces
to most of the elementary and special functions. Despite not being
a common special function, it has great potential as an analytical
tool in theoretical astrophysics, in particular to study the analytical
properties of density models such as the S\'ersic profile \citep{2011A&A...525A.136B,2011A&A...534A..69B}
and Einasto profile \citepalias{2012arXiv1202.5242R}. This function
will be included in future versions of the software \texttt{{Mathematica}}.
Additionally, several authors such as \citet{5426254} and \citet{ShafiqueAnsari2012}
have made available accurate and fast numerical routines to compute
the Fox $H$ function. Details about the many properties of this function
can be found in \citet*{0470263806}, \citet{srivastava1982h}, \citet{0415299160},
and \citet{mathai2009h}.

\noindent We remark that the shear $\gamma\left(x\right)$ is not
a directly measurable property owing to the mass-sheet degeneracy,
but that the measurable property is the \emph{reduced shear }\citep{1985ApJ...289L...1F,1988ApJ...327..693G,1995A&A...294..411S,1996astro.ph..6001N}

\noindent 
\begin{equation}
g\left(x\right)=\frac{\gamma\left(x\right)}{1-\kappa\left(x\right)}.\label{eq:reduced-shear}
\end{equation}

\subsection{First and second flexions \label{sec:Section3-2}}

Considering the weak lensing regime up to second order, two new lensing
properties can be recognized: the first flexion $\mathcal{F}$, which
describes the behaviour of the convergence gradient across the lensed
image and the second flexion $\mathcal{G}$, which describes the behaviour
of the shear gradient across the lensed image. These flexions produce
centre-shift and arc-like distortions that, with the addition of the
shear, cause the lensed image of the elliptical galactic source appear
to have a {}``banana-like'' shape.

\noindent The first flexion can be found by simply calculating the
convergence gradient, using eqs. (\ref{eq:convergence-Fox}) and (\ref{eq:1st-flex-pot-1}) 

\noindent \begin{flushleft}
\begin{equation}
\mathcal{F}\left(x\right)=\mathcal{F}_{0}\,\left\{ H_{2,3}^{2,1}\left[\begin{array}{c}
(-1,\,2),(0,\,1)\\
(0,\,2n),(-\frac{1}{2},\,1),(0,\,2)
\end{array}\biggr|\, x^{2}\right]\right\} e^{i\phi},\label{eq:flexion01-Fox}
\end{equation}

\par\end{flushleft}

\noindent with

\[
\mathcal{F}_{0}=\frac{\sqrt{\pi}\,\kappa_{{\text{c}}}\, D_{{\text{L}}}}{h\,\Gamma\left(n\right)}
\]

\noindent the flexion amplitude.

\noindent Combining eqs. (\ref{eq:2nd-flex-pot-1}) and (\ref{eq:shear-Fox}),
plus some algebra, the second flexion may be obtained

\noindent \begin{flushleft}
\begin{equation}
\mathcal{G}\left(x\right)=-\frac{\mathcal{F}_{0}}{2}\, G^{\prime}\left(x\right)\, e^{3i\phi},\label{eq:flexion02-Fox}
\end{equation}

\par\end{flushleft}

\noindent with

\begin{equation}
G^{\prime}\left(x\right)=H_{4,5}^{3,2}\left[\begin{array}{c}
(-1,\,2),(-\frac{1}{2},\,1),(0,\,1),(1,\,2)\\
(0,\,2n),(-\frac{1}{2},\,1),(2,\,2),(0,\,2),(-\frac{3}{2},\,1)
\end{array}\biggr|\, x^{2}\right].
\end{equation}

\noindent We have obtained analytical expressions for both flexions
expressed as Fox $H$ functions, where the flexions are circularly
symmetric as expected for the Einasto profile. The second flexion
with nine gamma functions in the integrand is a more complicated function
than the first flexion with only five gamma functions. 

\noindent As indicated before, the shear is affected by the so-called
mass-sheet degeneracy, and the same measurement difficulty arises and
with the first and second flexion. \citet{2008A&A...485..363S} demonstrated
that the observable properties are the \emph{reduced flexions}

\begin{equation}
F\left(x\right)=\frac{\mathcal{F}\left(x\right)+g\left(x\right)\,\mathcal{F}^{\star}\left(x\right)}{1-\kappa\left(x\right)}\label{eq:reduced-flexion01}
\end{equation}

\noindent and

\noindent 
\begin{equation}
G\left(x\right)=\frac{\mathcal{G}\left(x\right)+g\left(x\right)\,\mathcal{G}\left(x\right)}{1-\kappa\left(x\right)},\label{eq:reduced-flexion02}
\end{equation}

\noindent where $F\left(x\right)$ is a spin-0 field, $G\left(x\right)$
is a spin-3 field, and $\mathcal{F}^{\star}\left(x\right)$ is the
complex conjugate of the first flexion.

\noindent Given that the scope of this work is to present analytical
expressions of different weak lensing properties and the quantitative
comparison of these properties for several profiles, we focus on the
lensing properties given in eqs. (\ref{eq:shear-Fox}), (\ref{eq:flexion01-Fox}),
and (\ref{eq:flexion02-Fox}), instead of the reduced ones in eqs.
(\ref{eq:reduced-shear}), (\ref{eq:reduced-flexion01}), and (\ref{eq:reduced-flexion02}).

\subsection{Integer and half-integer values of $n$ \label{sec:Section3-3}}

We simplify the expressions for the shear in eq. (\ref{eq:shear-Fox})
and first, in eq. (\ref{eq:flexion01-Fox}), and second, in eq. (\ref{eq:flexion02-Fox}),
flexions in terms of the Fox $H$ function for rational values of
$n$, to the case when $n$ is an integer or half-integer number,
where the resulting expressions can be written in terms of the Meijer
$G$ function.

\noindent The Meijer $G$ function is defined by the Mellin-Barnes
integral \citep{1936NAvW.18..10}

\begin{multline}
G_{p,q}^{m,n}\left[\left.\begin{matrix}{\boldsymbol{a}}\\
{\boldsymbol{b}}
\end{matrix}\,\right|\, z\right]=\\
\frac{1}{2\pi i}\int_{{\cal \mathcal{L}}}\frac{\prod_{j=1}^{m}\Gamma(b_{j}+s)\prod_{j=1}^{n}\Gamma(1-a_{j}-s)}{\prod_{j=m+1}^{q}\Gamma(1-b_{j}-s)\prod_{j=n+1}^{p}\Gamma(a_{j}+s)}\, z^{-s}\,{\text{d}}s.\label{eq:defG}
\end{multline}

\noindent There is extensive literature about the Meijer $G$ function
and its many useful properties \citep{bateman1953higher,9780124599505,andrews1985special,prudnikov1990integrals}. 

\noindent For the Fox $H$ function, for there is not yet a numerical
implementation contrarily, there are various software packages with
Meijer $G$ numerical routines, such as the commercial \texttt{{Maple}},
\texttt{{Mathematica}}, and the free open-source \texttt{{Sage}}
and \texttt{{mpmath}} library.

\noindent By substituting the Gauss multiplication formula \citep{1970hmfw.book.....A}

\begin{equation}
\Gamma(2ny)=(2n)^{-\frac{1}{2}+2ny}\,(2\pi)^{\frac{1}{2}-n}\,\Gamma(y)\,\prod_{j=1}^{2n-1}\Gamma\left(\frac{j}{2n}+y\right),\label{eq:gauss-mult-form}
\end{equation}
into eq. (\ref{eq:Convergence-MellinBarnes}), and making use of eq.
(\ref{eq:defG}) for the comparison, we have \citepalias{2012arXiv1202.5242R}

\begin{subequations}
\label{Kappa-Meijer} 

\begin{equation}
\kappa\left(x\right)=\frac{\kappa_{\text{c}}}{2\,(2\pi)^{n-1}\,\sqrt{n}\,\Gamma\left(n\right)}\, x\, G_{0,2n}^{2n,0}\left[\begin{array}{c}
-\\
{\boldsymbol{b}}
\end{array}\biggr|\,\frac{x^{2}}{\left(2n\right)^{2n}}\right],
\end{equation}

\noindent where ${\boldsymbol{b}}$ is a vector of size $2n$ given
by

\noindent 
\begin{equation}
{\boldsymbol{b}}=\biggl\{\frac{1}{2n},\frac{2}{2n},\ldots,\frac{2n-1}{2n},-\frac{1}{2}\biggr\},\label{eq:vector-b}
\end{equation}

\end{subequations} 

\noindent which is an expression for the convergence of the Einasto
profile in terms of the Meijer $G$ function.

\noindent Now, substituting the convergence into eq. (\ref{eq:mean-kappa})
and performing the integration of Meijer $G$ function (eq. 07.34.21.0003.01
at the Wolfram Functions Site%
\footnote{http://functions.wolfram.com/HypergeometricFunctions/MeijerG/%
}), and inserting the integral product along with the convergence in
eq. (\ref{eq:shear}), we may write the shear as

\begin{subequations}
\label{shear-Meijer} 

\begin{multline}
\gamma\left(x\right)=\left\{ \frac{\kappa_{\text{c}}}{2\,(2\pi)^{n-1}\,\sqrt{n}\,\Gamma\left(n\right)}\, x\,\times\right.\\
\left.G_{1,2n+1}^{2n,1}\left[\begin{array}{c}
-\frac{1}{2}\\
{\boldsymbol{b}^{\prime}},-\frac{3}{2}
\end{array}\biggr|\,\frac{x^{2}}{\left(2n\right)^{2n}}\right]\right\} e^{2i\phi},
\end{multline}

\noindent where ${\boldsymbol{b}^{\prime}}$ is a vector of size $2n$
given by

\noindent 
\begin{equation}
{\boldsymbol{b}^{\prime}}=\biggl\{\frac{1}{2n},\frac{2}{2n},\ldots,\frac{2n-1}{2n},\frac{1}{2}\biggr\}.
\end{equation}

\end{subequations} 

\noindent Calculating the gradient for convergence following eq. (\ref{eq:1st-flex-pot-1})
and using the differentiation properties of Meijer $G$ function (eq.
07.34.20.0005.01  on the Wolfram Functions Site), we find

\begin{equation}
\mathcal{F}\left(x\right)=\left\{ \frac{\mathcal{F}_{0}}{(2\pi)^{n-1}\,\sqrt{n\,\pi}}\, G_{1,2n+1}^{2n,1}\left[\begin{array}{c}
-\frac{1}{2}\\
{\boldsymbol{b}},\frac{1}{2}
\end{array}\biggr|\,\frac{x^{2}}{\left(2n\right)^{2n}}\right]\right\} e^{i\phi}.\label{eq:flexion01-Meijer}
\end{equation}

\noindent Applying again the differentiation properties of Meijer
$G$ function to derive the gradient for shear according to eq. (\ref{eq:2nd-flex-pot-1}),
the result may be written as

\begin{multline}
\mathcal{G}\left(x\right)=\left\{ -\frac{\mathcal{F}_{0}}{(2\pi)^{n-1}\,\sqrt{n\,\pi}}\,\times\right.\\
\left.G_{3,2n+3}^{2n+1,2}\left[\begin{array}{c}
{-\frac{1}{2},-\frac{1}{2},\frac{1}{2}}\\
{\boldsymbol{b},\frac{3}{2},\frac{1}{2},-\frac{3}{2}}
\end{array}\biggr|\,\frac{x^{2}}{\left(2n\right)^{2n}}\right]\right\} e^{3i\phi}.\label{flexion02-Meijer}
\end{multline}

\noindent Additional simplifications in terms of generalized hypergeometric
function for half-integer values of $n$ can be found in Appendix
\ref{Appendix-A}.

\subsection{Simple cases: $n=1$ and $n=\frac{1}{2}$ \label{sec:Section3-3-1}}

For $n=1$, the density profile decreases exponentially from the system
centre

\begin{equation}
\rho\left(r\right)=\rho_{0}\,\exp\left\{ -\left(\frac{r}{h}\right)\,\right\} .\label{rho-n1}
\end{equation}

\noindent For the exponential case, the resulting weak lensing expressions
can be found by substituting $n=1$ in eqs. ~(\ref{Kappa-Meijer}),
~(\ref{shear-Meijer}), (\ref{eq:flexion01-Meijer}), and (\ref{flexion02-Meijer}) 

\begin{gather}
\kappa(x)=\frac{\kappa_{\text{c}}}{2}\, x\, G_{0,2}^{2,0}\left[\begin{array}{c}
-\\
\frac{1}{2},-\frac{1}{2}
\end{array}\biggr|\,\frac{x^{2}}{4}\right],\\
\left|\gamma\left(x\right)\right|=\frac{\kappa_{\text{c}}}{2}\, x\, G_{1,3}^{2,1}\left[\begin{array}{c}
-\frac{1}{2}\\
\frac{1}{2},\frac{1}{2},-\frac{3}{2}
\end{array}\biggr|\,\frac{x^{2}}{4}\right],\\
\left|\mathcal{F}\left(x\right)\right|=\frac{\mathcal{F}_{0}}{\sqrt{\pi}}\, G_{1,3}^{2,1}\left[\begin{array}{c}
-\frac{1}{2}\\
\frac{1}{2},-\frac{1}{2},\frac{1}{2}
\end{array}\biggr|\,\frac{x^{2}}{4}\right],\\
\left|\mathcal{G}\left(x\right)\right|=-\frac{\mathcal{F}_{0}}{\sqrt{\pi}}\, G_{3,5}^{3,2}\left[\begin{array}{c}
-\frac{1}{2},-\frac{1}{2},\frac{1}{2}\\
\frac{1}{2},-\frac{1}{2},\frac{3}{2},\frac{1}{2},-\frac{3}{2}
\end{array}\biggr|\,\frac{x^{2}}{4}\right].
\end{gather}

\noindent We may write the above Meijer $G$ functions in terms of 
Bessel functions 

\begin{gather}
\kappa\left(x\right)=\kappa_{\text{c}}\, x\, K_{1}\left(x\right)\label{kappa-n1}\\
\left|\gamma\left(x\right)\right|=\frac{4\,\kappa_{\text{c}}}{x^{2}}\,\left[1-\frac{x^{2}}{2}\, K_{2}\left(x\right)-\frac{x^{3}}{4}\, K_{1}\left(x\right)\right],\label{eq:shear-n1}\\
\left|\mathcal{F}\left(x\right)\right|=-\frac{\mathcal{F}_{0}}{\sqrt{\pi}}\, x\, K_{0}\left(x\right),\label{flexion01-n1}\\
\left|\mathcal{G}\left(x\right)\right|=\frac{\mathcal{F}_{0}}{\sqrt{\pi}}\,\left[\frac{16}{x^{3}}\left(1-\frac{x^{2}}{2}\, K_{2}\left(x\right)\right)-x\, K_{0}\left(x\right)-4\, K_{1}\left(x\right)\right],\label{eq:flexion02-n1}
\end{gather}

\noindent with $K_{\nu}(z)$ the modified Bessel function of the second
kind of order $\nu$.

\noindent For $n=\frac{1}{2}$, the Einasto mass distribution presents
a Gaussian fall-off

\begin{equation}
\rho\left(r\right)=\rho_{0}\,\exp\left\{ -\left(\frac{r}{h}\right)^{2}\,\right\} .\label{eq:pho-n-1/2}
\end{equation}

\noindent Therefore, we set $n=\frac{1}{2}$ in eqs. ~(\ref{Kappa-Meijer}),
~(\ref{shear-Meijer}), (\ref{eq:flexion01-Meijer}), and (\ref{flexion02-Meijer})
to obtain

\begin{gather}
\kappa(x)=\kappa_{\text{c}}\, x\, G_{0,1}^{1,0}\left[\begin{array}{c}
-\\
-\frac{1}{2}
\end{array}\biggr|\, x^{2}\right],\\
\left|\gamma\left(x\right)\right|=\kappa_{\text{c}}\, x\, G_{1,2}^{1,1}\left[\begin{array}{c}
-\frac{1}{2}\\
\frac{1}{2},-\frac{3}{2}
\end{array}\biggr|\, x^{2}\right],\\
\left|\mathcal{F}\left(x\right)\right|=2\,\mathcal{F}_{0}\, G_{1,2}^{1,1}\left[\begin{array}{c}
-\frac{1}{2}\\
-\frac{1}{2},\frac{1}{2}
\end{array}\biggr|\, x^{2}\right],\\
\left|\mathcal{G}\left(x\right)\right|=-2\,\mathcal{F}_{0}\, G_{3,4}^{2,2}\left[\begin{array}{c}
-\frac{1}{2},-\frac{1}{2},\frac{1}{2}\\
-\frac{1}{2},\frac{3}{2},\frac{1}{2},-\frac{3}{2}
\end{array}\biggr|\, x^{2}\right].
\end{gather}

\noindent We can equivalently express these Meijer $G$ functions
in terms of elementary functions 
\begin{gather}
\kappa\left(x\right)=\kappa_{\text{c}}\,{\text{e}}^{-x^{2}},\label{kappa-n-1/2}\\
\left|\gamma\left(x\right)\right|=\frac{\kappa_{\text{c}}}{x^{2}}\,{\text{e}}^{-x^{2}}\,\left({\text{e}}^{x^{2}}-1-x^{2}\right),\label{eq:shear-n-1/2}\\
\left|\mathcal{F}\left(x\right)\right|=-2\,\mathcal{F}_{0}\, x\,{\text{e}}^{-x^{2}},\label{flexion01-n-1/2}\\
\left|\mathcal{G}\left(x\right)\right|=\frac{4\mathcal{\, F}_{0}}{x^{3}}\,{\text{e}}^{-x^{2}}\,\left[{\text{e}}^{x^{2}}-1-\frac{x^{2}}{2}\left(x^{2}+2\right)\right].\label{flexion02-n-1/2}
\end{gather}

\noindent The results for the exponential profile described in eqs.
(\ref{kappa-n1}), (\ref{eq:shear-n1}), (\ref{flexion01-n1}), and
(\ref{eq:flexion02-n1}), and for the Gaussian profile in eqs. (\ref{kappa-n-1/2}),
(\ref{eq:shear-n-1/2}), (\ref{flexion01-n-1/2}), and (\ref{flexion02-n-1/2})
can be checked separately by substituting the density profiles in eqs.
(\ref{rho-n1}) and (\ref{eq:pho-n-1/2}) into eq. (\ref{eq:Convergence_integral})
and performing the relevant derivations for the shear and flexions.

\subsection{Asymptotic behaviour \label{sec:Section3-4}}

The behaviour of the weak lensing properties of the Einasto profile
at small radii ($x\ll1$) can be studied by using the series expansions
of Appendix \ref{Appendix-B}. We easily arrive at appropriate expressions,
which depend on the value of $n$ for the shear

\begin{align}
\left|\mathcal{\gamma}\left(x\right)\right| & \sim\frac{\kappa_{\text{c}}}{4}\,\frac{\Gamma\left(1-n\right)}{\Gamma\left(1+n\right)}\, x^{2}\qquad &  & \text{for \ensuremath{n<1}},\\[3mm]
\left|\mathcal{\gamma}\left(x\right)\right| & \sim\frac{\kappa_{\text{c}}}{4}\, x^{2}\,\left[-\ln\left(\frac{x}{2}\right)+\frac{1}{4}-\gamma\right]\qquad &  & \text{for \ensuremath{n=1}},\\
\left|\mathcal{\gamma}\left(x\right)\right| & \sim\frac{\kappa_{\text{c}}\,\sqrt{\pi}}{2n\,\left(3n+1\right)}\,\left[\frac{\Gamma\left(\frac{n-1}{2n}\right)}{\Gamma\left(n\right)\,\Gamma\left(\frac{2n-1}{2n}\right)}\right]\, x^{1+1/n} &  & \text{for \ensuremath{n>1.}}
\end{align}

\noindent The first flexion near zero behaves as 

\begin{align}
\left|\mathcal{F}\left(x\right)\right| & \sim-\frac{\mathcal{F}_{0}}{n\,\sqrt{\pi}}\,\Gamma\left(1-n\right)\, x\qquad &  & \text{for \ensuremath{n<1}},\\[3mm]
\left|\mathcal{F}\left(x\right)\right| & \sim\frac{\mathcal{F}_{0}}{n\,\sqrt{\pi}}\, x\,\left[\ln\left(\frac{x}{2}\right)+1+\gamma\right]\qquad &  & \text{for \ensuremath{n=1}},\\
\left|\mathcal{F}\left(x\right)\right| & \sim-\frac{\mathcal{F}_{0}}{2n^{2}}\,\left[\frac{\Gamma\left(\frac{n-1}{2n}\right)}{\Gamma\left(\frac{2n-1}{2n}\right)}\right]\, x^{1/n}\qquad &  & \text{for \ensuremath{n>1.}}
\end{align}

\noindent For the second flexion, the asymptotic behaviour in the
neighbourhood of the lensed image origin is described by

\begin{align}
\left|\mathcal{G}\left(x\right)\right| & \sim-\frac{\mathcal{F}_{0}}{6n\,\sqrt{\pi}}\,\Gamma\left(1-3n\right)\, x^{3}\qquad &  & \text{\text{for \ensuremath{n<\frac{1}{3}}\ or \ensuremath{n=\frac{1}{2}}}},
\end{align}
\begin{align}
\left|\mathcal{G}\left(x\right)\right| & \sim-\frac{9\,\mathcal{F}_{0}}{32\,\sqrt{\pi}}\,\Gamma\left(\frac{1}{15}\right)\, x^{5}\quad\qquad\qquad\qquad\nonumber \\
 & +\frac{\mathcal{F}_{0}}{\sqrt{\pi}}\, x^{3}\,\left[\frac{1}{3}\ln\left(\frac{x}{2}\right)+\frac{5}{2}+\gamma\right] &  & \text{for \ensuremath{n=\frac{1}{3}}},
\end{align}


\begin{align}
\left|\mathcal{G}\left(x\right)\right|\sim-\frac{3\mathcal{F}_{0}}{32n\,\sqrt{\pi}}\,\Gamma\left(1-5n\right)\, x^{5}\qquad\nonumber \quad\qquad\quad \\ 
+\frac{\mathcal{F}_{0}}{2n^{2}}\,\left(\frac{n^{2}-1}{3n+1}\right)\,\left[\frac{\Gamma\left(-\frac{n+1}{2n}\right)}{\Gamma\left(-\frac{1}{2n}\right)}\right]\, x^{1/n} &  & \text{ for \ensuremath{n>\frac{1}{3}}},
\end{align}

\begin{align}
\left|\mathcal{G}\left(x\right)\right| & \sim\frac{\mathcal{F}_{0}}{\sqrt{\pi}}\,\frac{x^{3}}{12}\,\left[\ln\left(\frac{x}{2}\right)-\frac{2}{3}+\gamma\right]\qquad &  & \text{for \ensuremath{n=1}},\\
\left|\mathcal{G}\left(x\right)\right| & \sim\frac{\mathcal{F}_{0}}{2n^{2}}\left(\frac{n^{2}-1}{3n+1}\right)\,\left[\frac{\Gamma\left(-\frac{n+1}{2n}\right)}{\Gamma\left(-\frac{1}{2n}\right)}\right]\, x^{1/n}\qquad &  & \text{for \ensuremath{n>1.}}
\end{align}

\noindent The power-law and logarithmic series converge very slowly
at large radii ($x\gg1$). Hence, we cannot use them to investigate
the behaviour of the shear and flexions at large radii. However, following
\citet{KilbasSaigo99}, we derive asymptotic expansions at large radii
for these properties. For the shear, we have 

\noindent 
\begin{multline}
\left|\mathcal{\gamma}\left(x\right)\right|\sim2\,\frac{\Gamma\left(3n\right)}{\Gamma\left(n\right)}\,\kappa_{\text{c}}\, x^{-2}-\frac{\sqrt{2\,\pi}}{\sqrt{n}\,\Gamma\left(n\right)}\,\kappa_{\text{c}}\,{\text{e}}^{-x^{1/n}}\, x^{1-\frac{1}{2n}}.\label{eq:shear-larga}
\end{multline}

\noindent When $x\rightarrow\infty$, the first flexion behaves as

\begin{equation}
\left|\mathcal{F}\left(x\right)\right|\sim-\frac{\sqrt{2}\,\mathcal{F}_{0}}{n^{3/2}}\,{\text{e}}^{-x^{1/n}}\, x^{\frac{1}{2n}},
\end{equation}

\noindent and the second flexion is characterised by the behaviour

\begin{equation}
\left|\mathcal{G}\left(x\right)\right|\sim\frac{\sqrt{2}\,\mathcal{F}_{0}}{n^{3/2}}\,{\text{e}}^{-x^{1/n}}\, x^{\frac{1}{2n}}.
\end{equation}

\section{Profile comparisons \label{sec:Section4}}

We compare the weak lensing properties for the Einasto profile obtained
in the Section \ref{sec:Section3} with the properties of the singular
isothermal sphere, the Navarro-Frenk-White profile, and S\'ersic
model. The properties for these models can be found in \citealp{1996A&A...313..697B},
\citealp{2000ApJ...534...34W}, \citetalias{2006MNRAS.365..414B},
and \citealp{2009MNRAS.396.2257L}. We follow the same approach that
was used by \citet{2000ApJ...534...34W} and \citet{2009MNRAS.396.2257L},
which consists in fixing the halo mass $M_{200}$ and permits the
calculation of the virial radius. We use $M_{200}=1\times10^{12}\, h^{-1}\, M_{\odot}$
and $z_{l}=0.4$, which are the approximate average galactic-sized
halo mass and lens redshift, respectively, found by \citet{2007ApJ...669...21P}
in their galaxy-galaxy weak lensing analysis of the Canada-France-Hawaii
Telescope Legacy Survey (CFHTLS). We choose to place the source at
$z_{s}=0.92$, which implies $D_{\text{LS}}/D_{\text{S}}\simeq0.5$.
We assume that the concentration-mass varies with the halo redshift
in the range $0\leq z\leq2$ as \citep{2008MNRAS.390L..64D}

\begin{equation}
c_{200}\left(M_{200},z\right)={\text{A}}\,\left(\frac{M_{200}}{M_{pivot}}\right)^{{\text{B}}}\left(1+z\right)^{{\text{C}}},\label{eq:c200}
\end{equation}

\noindent where $M_{pivot}=2\times10^{12}\, h^{-1}\, M_{\odot}$ and
$\left\{ \,{\text{A}},\,{\text{B}},\,{\text{C}}\,\right\} =\left\{ \,6.40,-0.108,-0.62\,\right\} $
for the Einasto profile. Using the above relation, we find a concentration
of $c_{{\text{E}}}=5.80$. In the case of the Einasto halo, we alse
choose to use a value of $n\simeq6$ that corresponds to the halo
mass that is used according to \citet{2008MNRAS.387..536G}. For the
NFW halo, the concentration can be determined using eq. (\ref{eq:c200})
with $\left\{ \,{\text{A}},\,{\text{B}},\,{\text{C}}\,\right\} =\left\{ \,5.71,-0.048,-0.47\,\right\} $
\citep{2008MNRAS.390L..64D}, we obtain $c_{{\text{NFW}}}=5.31$.
The calculation of the parameters model is explicit, except in the
case of the S\'ersic profile, for which we construct the model using
the procedure outlined in Appendix B of \citet{2009MNRAS.396.2257L}.
This procedure requires employing the \citet{1997A&A...321..111P}
model, which is an analytical approximation of the S\'ersic deprojection;
we use a S\'ersic index of $m\simeq8.6$.

\begin{figure*}
\includegraphics[width=1\columnwidth]{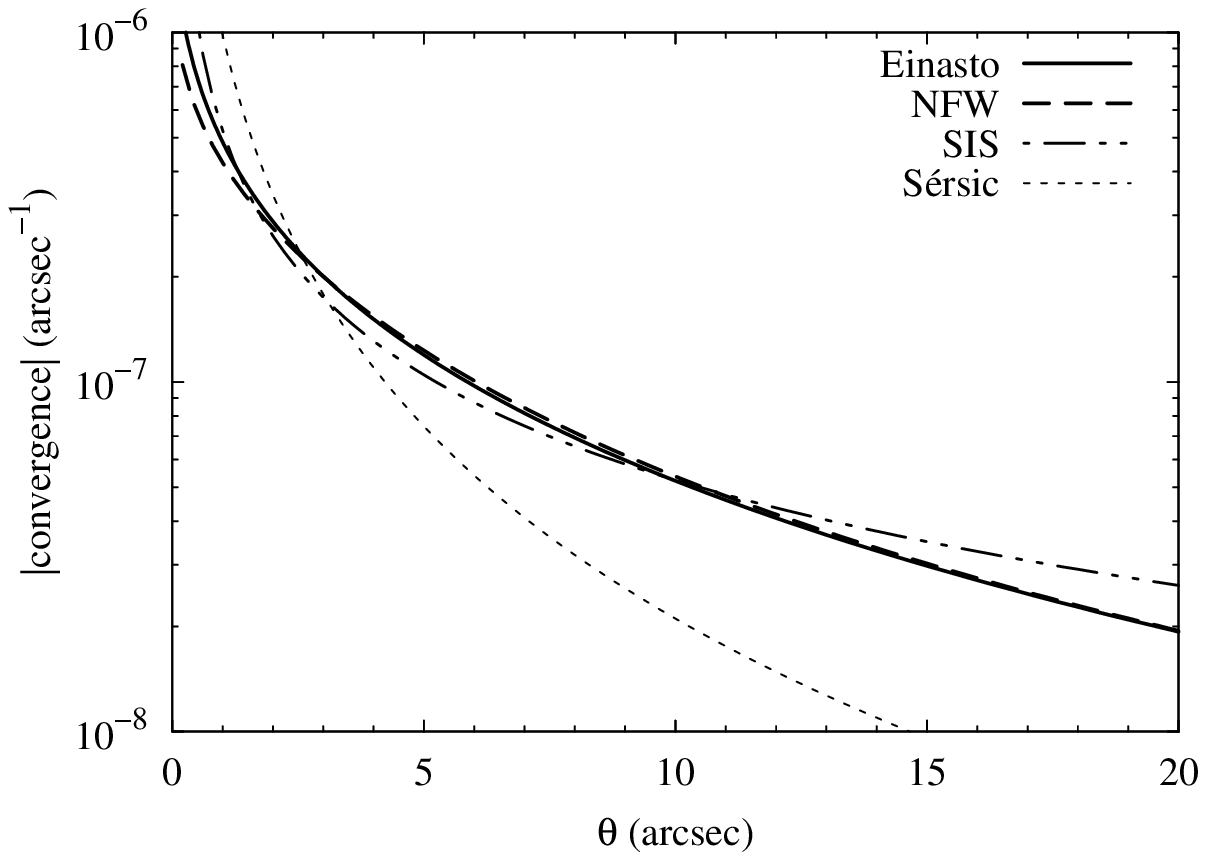}\includegraphics[width=1\columnwidth]{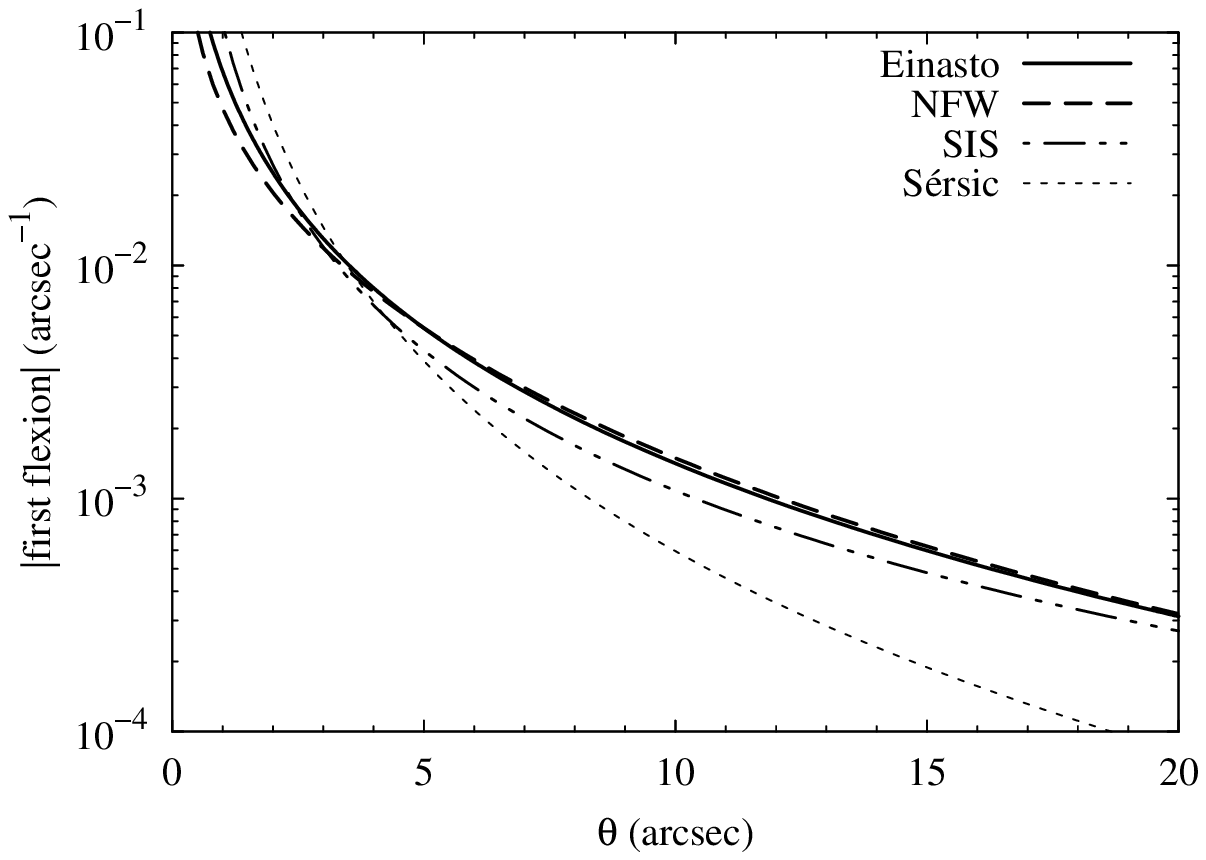}

\includegraphics[width=1\columnwidth]{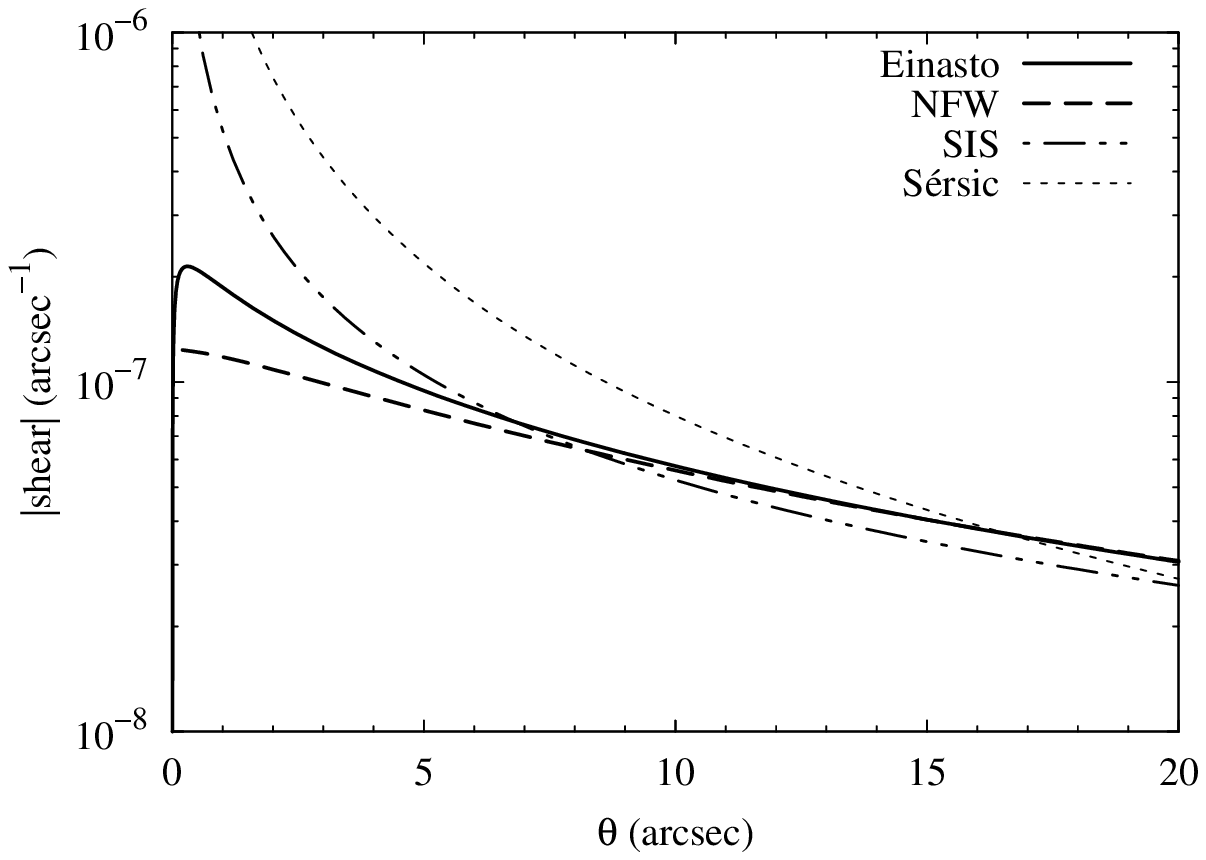}\includegraphics[width=1\columnwidth]{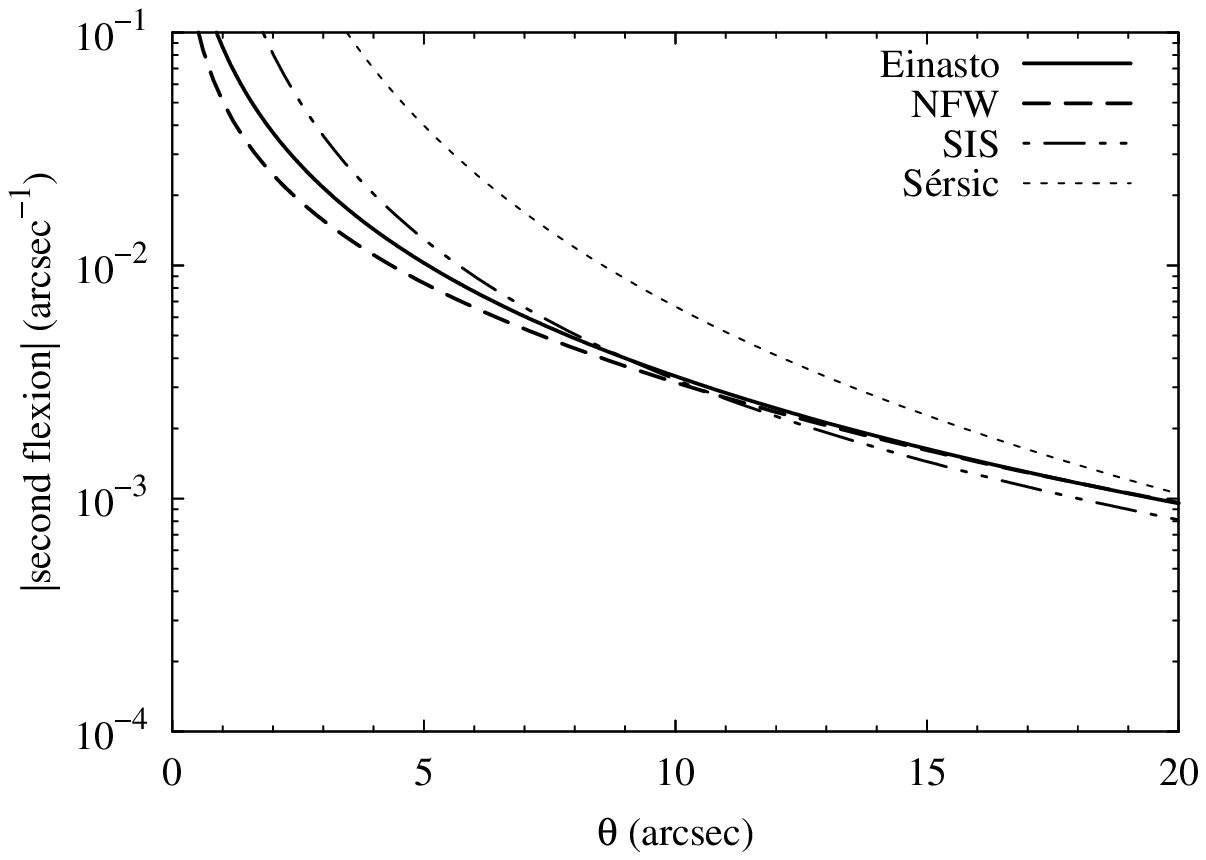}\caption{Comparison of the magnitude of convergence, shear, first and second
flexion due to a dark matter halo of mass $M_{200}=1\times10^{12}\, h^{-1}\, M_{\odot}$,
assuming that the mass distribution is described by Einasto, NFW,
and SIS profiles and the S\'ersic model. The lens and source planes
are located at redshifts $z_{\text{L}}=0.4$ and $z_{\text{S}}=0.92$,
respectively. The Einsten radius for the SIS is $\theta_{E}=0.216^{\prime\prime}$,
indicating the angular distance when the weak lensing effects are
relevant.}

\label{comparison} 
\end{figure*}

\noindent Figure \ref{comparison} shows the convergence, shear, and
both first and second flexion expected for a galatic-sized halo, assuming that
the mass distribution is given by the Einasto, SIS, NFW, and S\'ersic
profiles as a function of the angular separation from the lens centre.
An indicator of where the weak lensing effects are relevant is the
Einstein radius of the SIS profile, which is $\theta_{E}=0.216^{^{\prime\prime}}$
for the halo that is being studied. As can be seen, the overall behaviours
of the profiles are comparable, the differences between the magnitudes
of the lensing quantities being stronger at smaller angular separations,
with the shear and second flexion showing a greater dissimilarity
than the convergence and first flexion. These dissimilarities in the
central region indicate that the Einasto profile may be differentiable
from the other profiles in observational weak lensing studies. Our
result agrees with \citet{2010A&A...520A..30M}, who compared the
shear for the Einasto and NFW profiles finding that the major differences
between the profiles are at small distances. As can be seen, the magnitudes
of all properties at small distances are stronger for the SIS than
the NFW profile and large distances the relation is inverted. In
addition, the magnitudes are stronger for the NFW profile than that
of SIS, which is consistent with the slope for the SIS being $-2$,
and for the NFW profile the inner slope being $-1$ and the outer
being $-3$. Interestingly, we also note that the lensing properties
of the Einasto profile tend to be very similar to those of the NFW
profile for increasing angular separation. 

\begin{figure*}
\includegraphics[width=1\columnwidth]{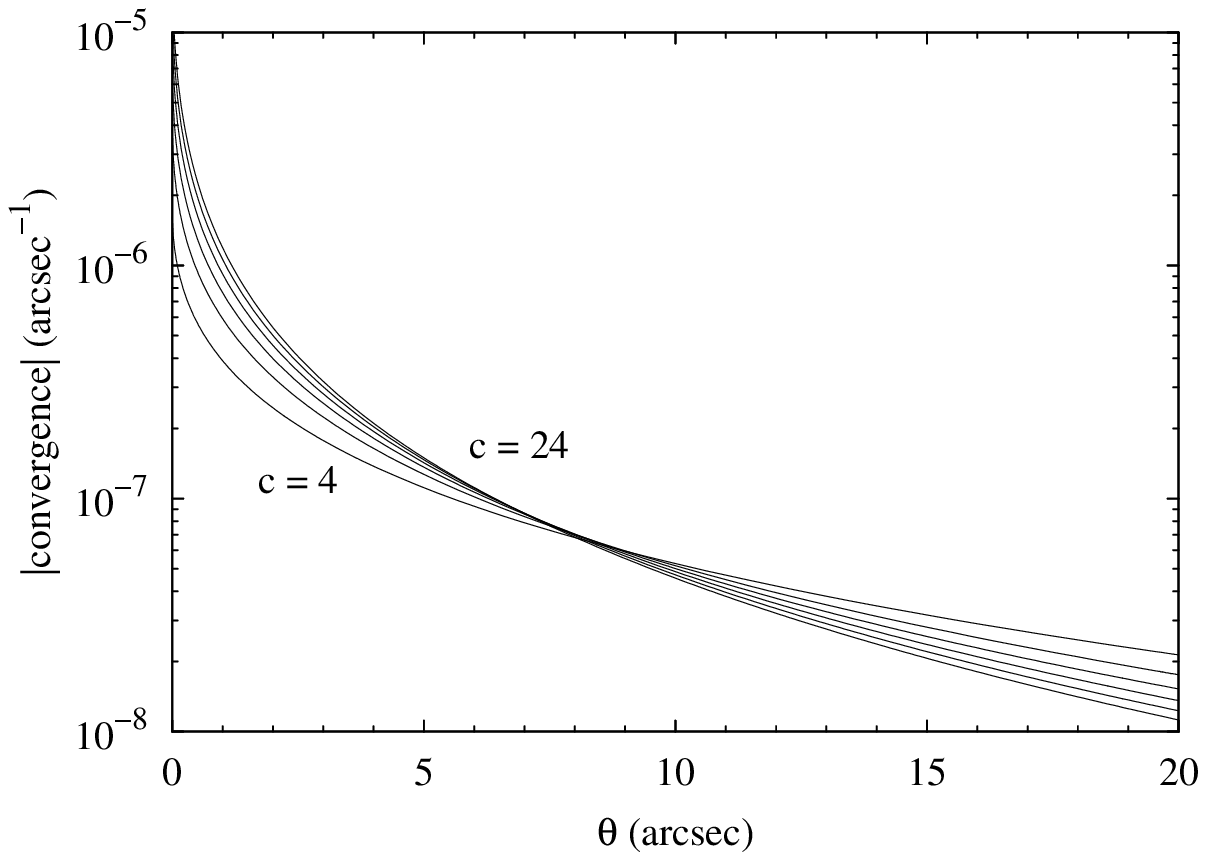}\includegraphics[width=1\columnwidth]{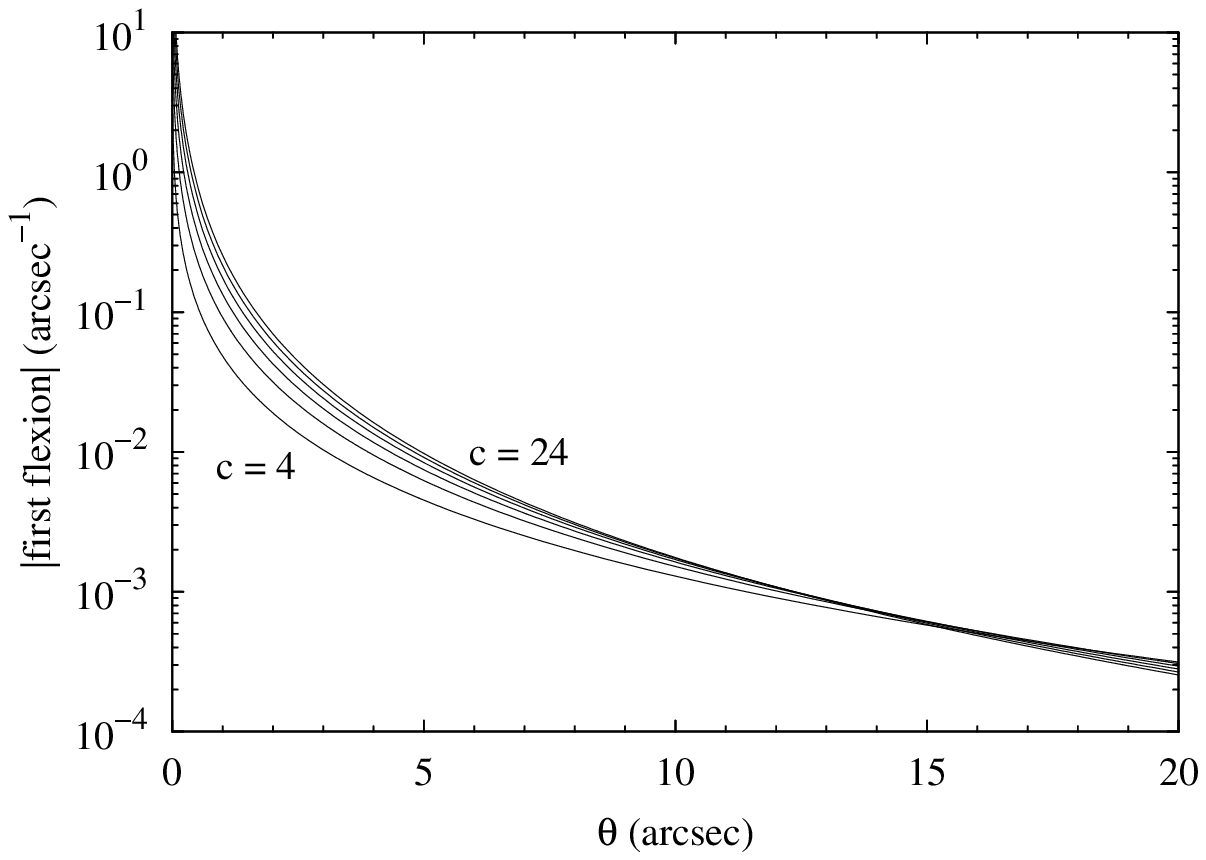}

\includegraphics[width=1\columnwidth]{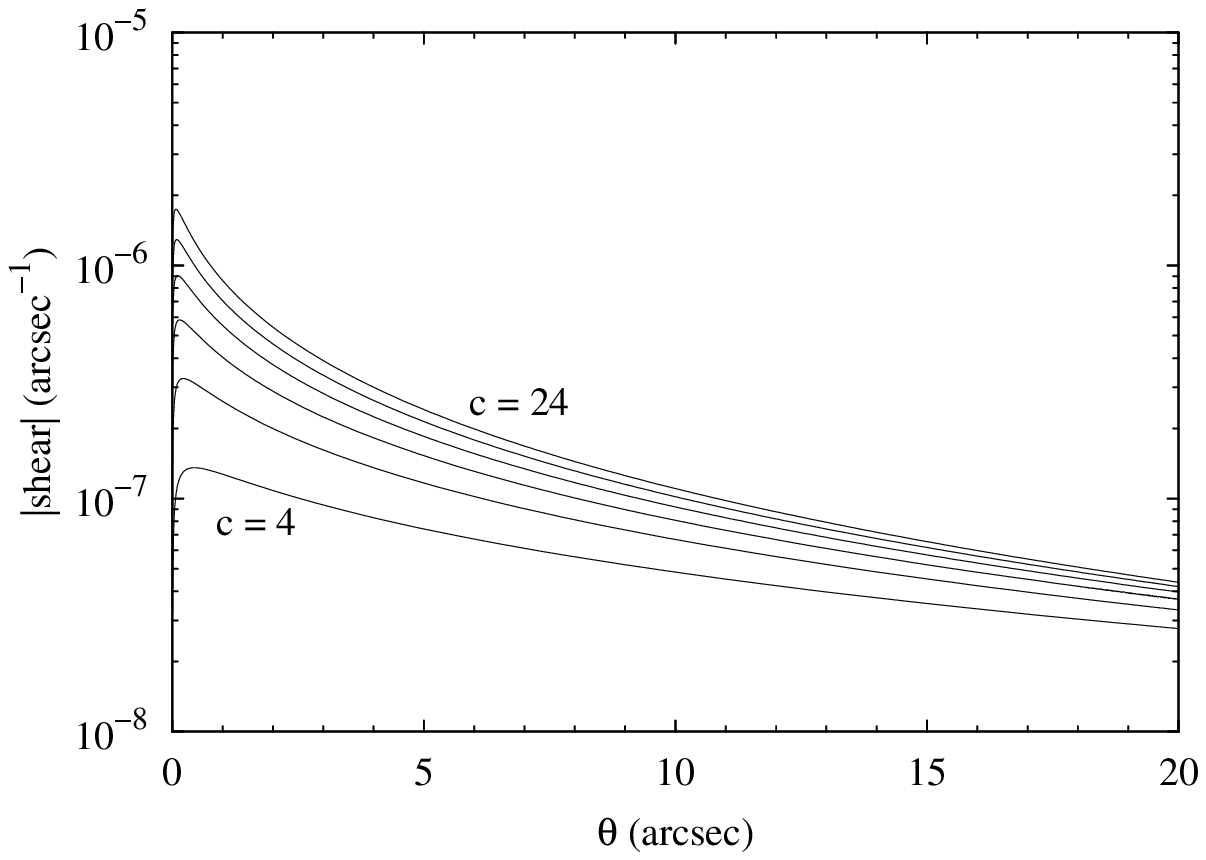}\includegraphics[width=1\columnwidth]{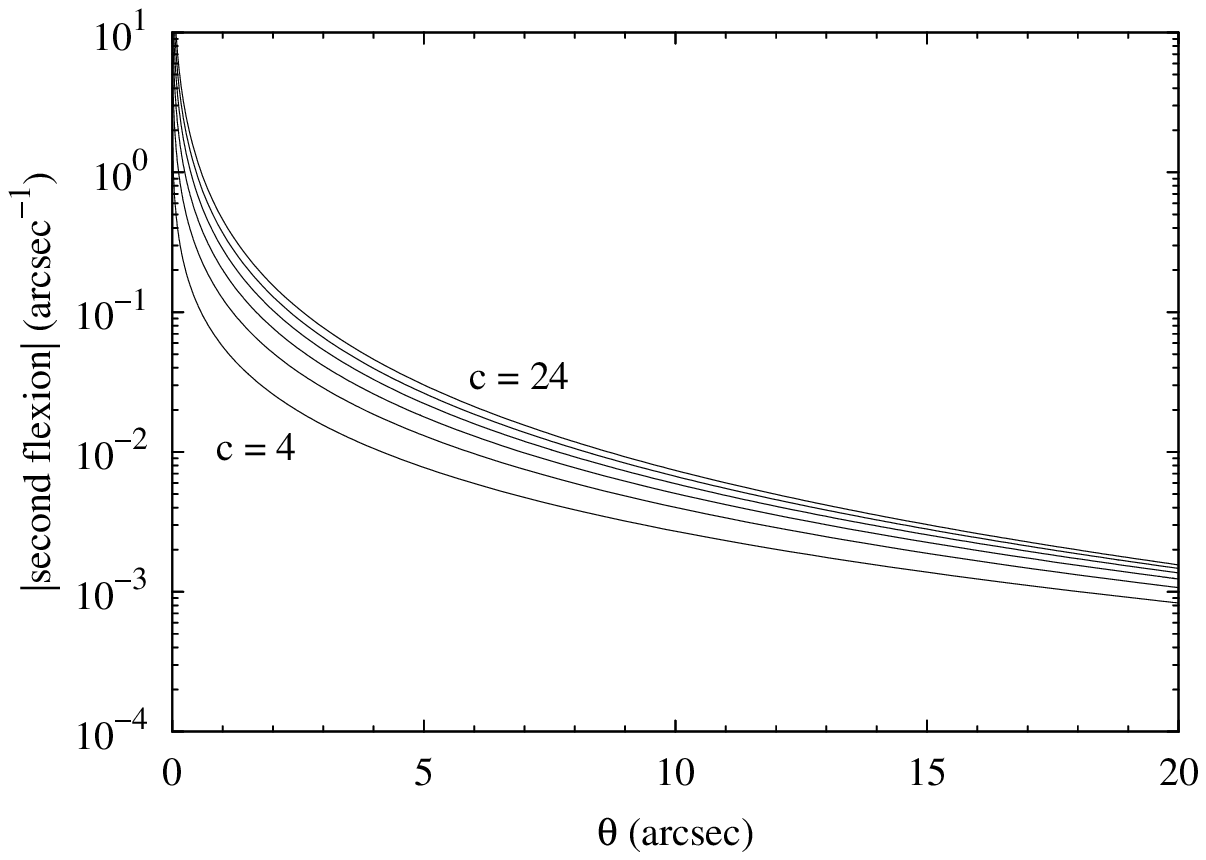}\caption{Comparison of the magnitude of convergence, shear, first and second
flexion due to an Einasto dark matter halo of mass $M_{200}=1\times10^{12}\, h^{-1}\, M_{\odot}$,
for different values of the concentration $c=4$, $8$, $12$, $16$,
$20$, $24$.}

\label{convergence} 
\end{figure*}

\noindent In Figure \ref{convergence}, we plot dependence of the
weak lensing properties on the concentration for the Einasto profile.
We use the same mass as in the previous comparison, and values of
$c=4$, $8$, $12$, $16$, $20$, $24$ for the concentration. The
concentration dependence of these properties is clearly non-linear,
and very similar to that of the NFW profile investigated
by \citet{2009MNRAS.396.2257L}. As with the NFW profile, the concentration
dependence is far stronger at small separations. With increasing angular
distance from the lensed image the dependence gets less pronounced
and the curves become almost identical, except the convergence,
which curves seem to be distinguishable at both small and large angular
distances. The shear and second flexion seem to be more sensitive
to the concentration variation than the convergence and first flexion,
making the first two lensing properties useful tools for breaking the
degeneracy between mass and concentration, if this occurs, as discussed
by \citet{2009MNRAS.396.2257L} for the NFW profile. 

\begin{figure*}
\includegraphics[width=1\columnwidth]{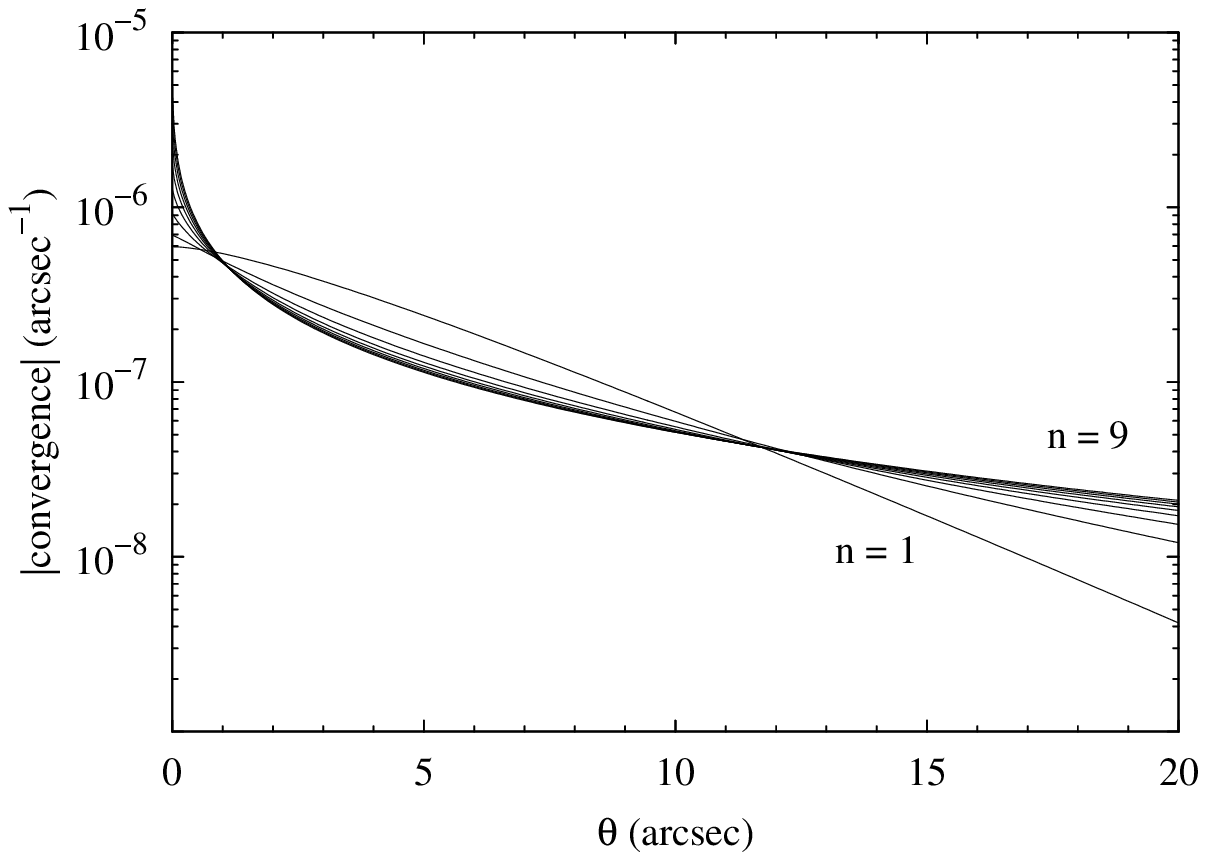}\includegraphics[width=1\columnwidth]{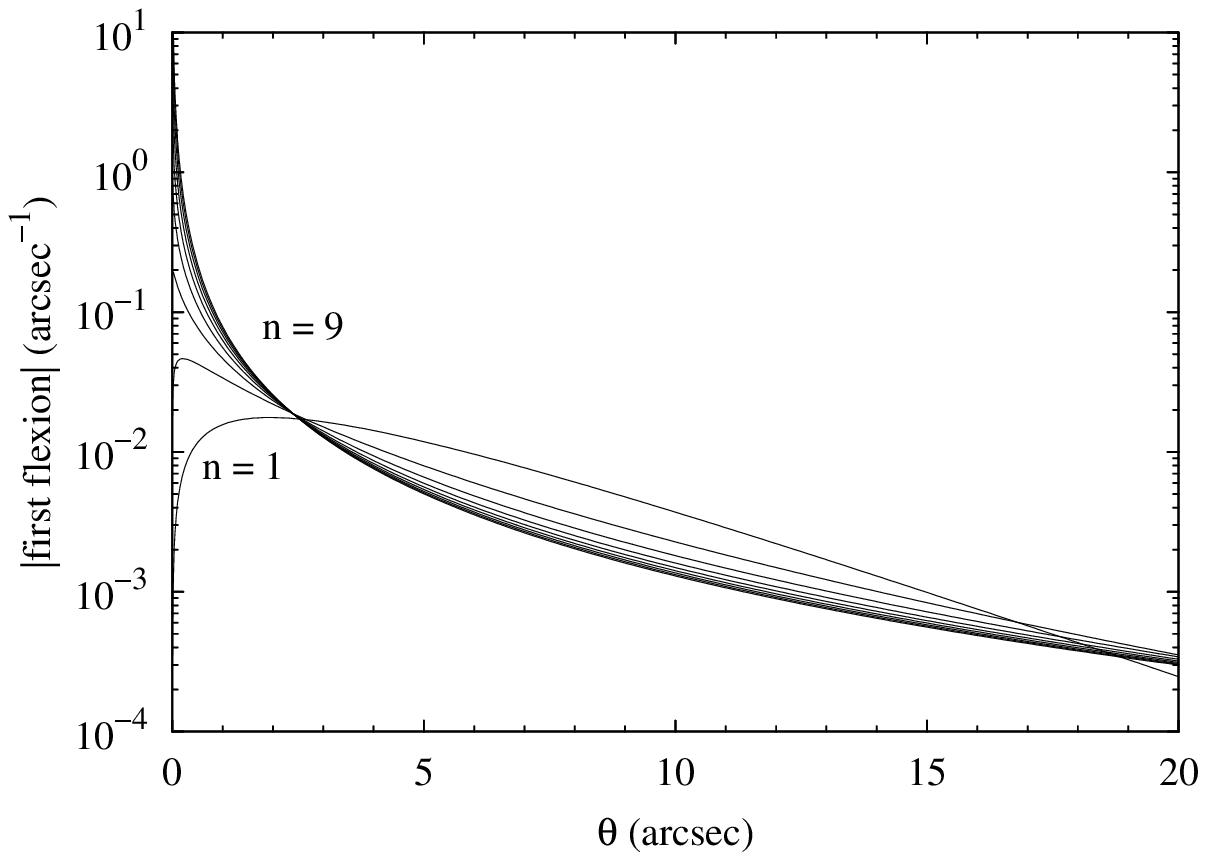}

\includegraphics[width=1\columnwidth]{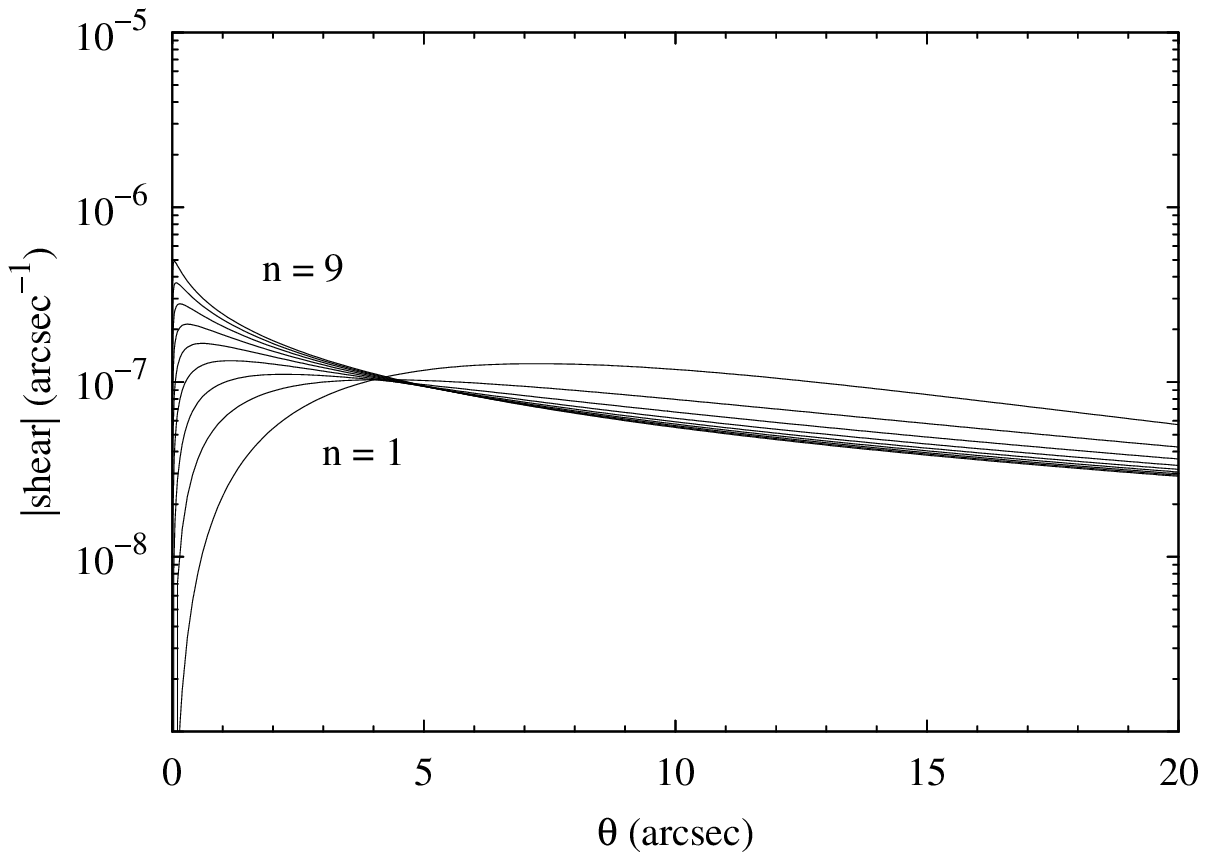}\includegraphics[width=1\columnwidth]{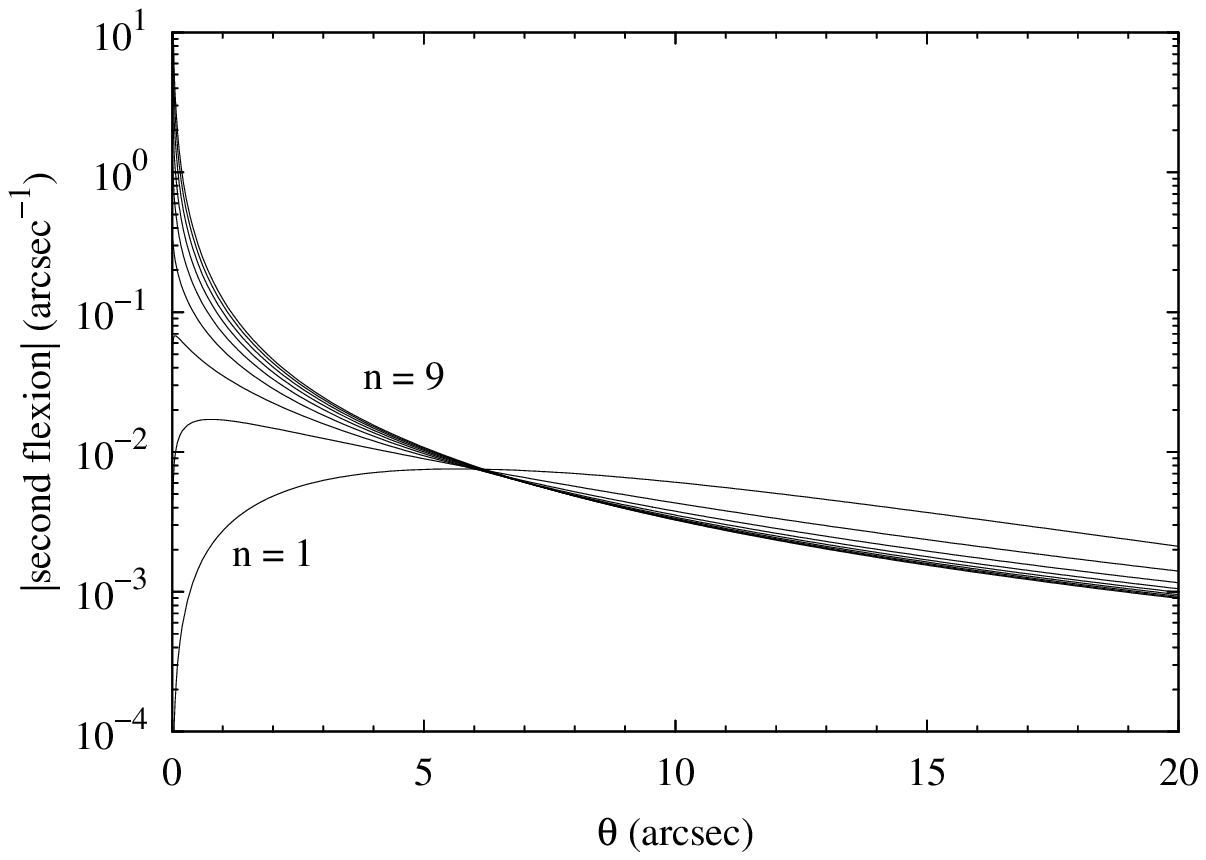}\caption{Comparison of the magnitude of convergence, shear, first and second
flexion due to an Einasto dark matter halo of mass $M_{200}=1\times10^{12}\, h^{-1}\, M_{\odot}$,
for different values of the Einasto index, $n$, between $1\leq n\leq9$.}

\label{Einasto-indexes} 
\end{figure*}

\noindent We illustrate the index dependence of the Einasto profile
in Figure \ref{Einasto-indexes}. We again use the same halo mass
$M_{200}=1\times10^{12}\, h^{-1}\, M_{\odot}$, and vary the Einasto
index, $n$, across the range $1\leq n\leq9$. It is evident, that
the index dependence is greater at smaller angular separations for
the lensing properties, becoming marginally distinguishable at increasing
$\theta$, with the only two exceptions being the exponential profile
($n=1$), which can be clearly identified, and with a little bit more
work the $n=2$ profile can be distinguished as well. The convergence
is the less dependent on $n$ of all the lensing quantities; the shear
and the first and second flexions depend more strongly on
$n$ than the convergence, near the lensed halo centre; this feature
is very important, because it is near the lensed image that the lensing
signal is stronger, making the first flexion, shear, and second flexion
excellent tools for constraining $n$, whilst the convergence may
be used to derive the halo mass. In contrast to the concentration
dependence of the Einasto profile, which is similar to that of the
NFW profile, the index dependence of the Einasto profile differs from
the corresponding dependence of the S\'ersic profile. \citet{2009MNRAS.396.2257L}
found that for the S\'ersic profile the index dependence is stronger
in $\kappa$ and $\mathcal{F}$, and very weak in $\gamma$ and $\mathcal{G}$,
whenever this dependence is present for the S\'ersic profile, its
effect decreases with boosted angular separations, where the lensing
signal is weaker, making it difficult to constraint the S\'ersic
index. Thereby, we conclude that may be easier to constrain the index
for an Einasto halo rather that for a S\'ersic halo. 

\section{Summary and conclusions \label{sec:Section5}}

We have applied the Mellin transform technique to obtain closed-form
expressions for the weak lensing properties of the Einasto profile.
The expressions for the shear $\gamma\left(x\right)$, first flexion
$\mathcal{F}\left(x\right)$, and second flexion $\mathcal{G}\left(x\right)$
can be written in terms of the Fox $H$ function for general values
of the Einasto index $n$, and can simplified in terms of the Meijer
$G$ function for integer or half-integer values of $n$. We utilized
the residue theorem to calculate specific power and logarithmic-power
expansions for these expressions. The expansions permit us to study
the asymptotic behaviour of the weak lensing properties at small and
large radii. Furthermore, we employed the Slater's theorem \citep{marichev1983handbook}
to derive an expression for the convergence $\kappa\left(x\right)$
in terms of the generalized hypergeometric function, which is valid
for half-integer values of $n$. This enables the other expressions
for the lensing properties to be written in terms of the hypergeometric
function. 

\noindent We have examined in detail the convergence, shear, and first
and second flexions for an Einasto profile and other profiles including
the singular isothermal sphere, the NFW, and S\'ersic profiles. We
found that the Einasto profile overall has a similar behaviour to
these profiles. Nonetheless, this profile is clearly different from
the others, particularly at small angular separations from the lens
centre, where the lensing signal is stronger. At large angular separations,
the Einasto profile behaves far very similarly to the NFW profile than the
other profiles. We explored the dependence of the Einasto profile
on the concentration parameter, our results indicating that it has
a non-linear concentration dependence and that the shear and second
flexion are more effective indicators of the dependence than the convergence
and the first flexion. In addition, we studied the Einasto index dependence.
For this parameter, the dependence seems to be weaker than for the
concentration, for which the shear, second, and first flexions seem
to be more sensitive to the index dependence that the convergence.
We note that the magnitude of the lensing properties of a S\'ersic
model are stronger than for the other profiles, this indicates that
the profile selected to model the halo must be chosen with caution
as discussed by \citet{2009MNRAS.396.2257L}. We note that the index
dependence of an Einasto halo is stronger at small angular distances
from the lens centre, which is the opposite of the case for a S\'ersic
halo for which at large angular distances the dependence is stronger.
This means that observationally it is easier to constrain the value
of the index for the Einasto profile rather than the S\'ersic model,
because the lensing signal is stronger near the lensed image.

\noindent The availability of analytical expressions for the Einasto-profile
weak-lensing properties is of foremost importance, and constitutes
an effort to foster the inclusion of this density profile in weak
lensing modelling studies. There are several possible applications
of our results for modelling studies. For example, one of them is
the generation of the shear signal in weak lensing analyses, which
can provide valuable information about the description of the mass
density profiles. This, in turn, places constraints on the model parameters
such as mass, concentration, and particularly the Einasto index, which
is known to scale with mass and redshift according to $N$-body simulations
\citep{2008MNRAS.387..536G,2008MNRAS.388....2H}, and for which our
results could be used to verify this variation observationally. Likewise,
weak flexion might be generated using our expressions and used to
constrain the model parameters, particularly when halo substructure
has to be proven, providing another scale within the haloes, where
the behaviour of $n$ can be studied.

\noindent Here, we considered spherically symmetrical haloes, but
haloes are far from ideal symmetric objects (see e.g. \citet{2006ApJ...646..815S,2007MNRAS.376..215B,2010MNRAS.407..891H}).
Nevertheless, weak shear has proven successfully in studying deviations
in the halo shape from spherical symmetry, such as the halo ellipticity
\citep{2004ApJ...606...67H,2006MNRAS.370.1008M,2007ApJ...669...21P,2009ApJ...695.1446E,2010MNRAS.405.2215O}.
Similarly, weak flexion has been proposed as a tool to investigate
the halo ellipticity \citep{2009MNRAS.400.1132H,2011A&A...528A..52E,2011MNRAS.417.2197E}.
Triaxiality is another aspect of the halo shape that has been explored
using weak lensing \citep{2005ApJ...632..841O,2005A&A...443..793G,2011MNRAS.416.3187S,2012MNRAS.420..596F};
ignoring the halo triaxiality can affect the parameter estimation
in lens-rich clusters \citep{2007MNRAS.380..149C,2009MNRAS.393.1235C},
leading to the cluster appearing to be more massive and concentrated,
particularly, if its major axis is aligned with the line of sight
\citep{0004-637X-654-2-714,MNR:MNR14154,A&A2010...Meneghetti}. Some
cluster studies where there are apparently lensing biases in the estimated
concentration and mass include that of \citet{2008ApJ...685L...9B},
who analysed the mass and concentration of four nearly relaxed clusters,
that of the gravitational lens with the largest Einstein radius detected
so far MACS J0717.5+3745 \citep{2009ApJ...707L.102Z}, and \citet{2009ApJ...699.1038O},
who obtained the radial profile of four clusters combining lensing
data from the Subaru telescope. The inclusion of our results in weak
lensing ellipticity and triaxiality studies is straightforward, therefore
it enables the possibility of investigating the halo ellipticity and
triaxiality with the Einasto profile.

\noindent Our current knowledge of the structure of the Universe on
large scales will be improved by new gravitational lensing surveys
such as the Dark Energy Survey%
\footnote{http://www.darkenergysurvey.org/%
} (DES), Euclid%
\footnote{http://sci.esa.int/euclid/%
}, the Large Synoptic Survey Telescope (LSST)%
\footnote{http://www.lsst.org/lsst/%
}, the James Webb Space Telescope%
\footnote{http://www.jwst.nasa.gov/%
}, KiDS%
\footnote{http://kids.strw.leidenuniv.nl/%
}, Pan-STARRS%
\footnote{http://pan-starrs.ifa.hawaii.edu/public/%
}, and WFIRST%
\footnote{http://wfirst.gsfc.nasa.gov/%
}. These surveys will provide more accurate measurements of weak lensing
that could be modelled using the analytical expressions presented
in this work.

\noindent This paper constitutes a further step in studying the properties
of the Einasto profile using analytical means. In addition, it extends
and complements the work of \citetalias{2012arXiv1202.5242R}, providing
additional simplified expressions for their results. With this work,
we hope to encourage the use of special functions such as the Fox
$H$ function, the Meijer $G$ function, and the generalized hypergeometric
function in astronomy and astrophysics.

\noindent \begin{acknowledgements} ERM and FFA wish to thank H. Morales, R. Carboni, J. Gutirrez, R. Maga\~{n}a, and M. Chaves for their critical reading of the manuscript. Moreover, we wish to thank the referee for valuable comments and suggestions. This research has made use of NASA's Astrophysics Data System Bibliographic Services. \end{acknowledgements} 

\bibliographystyle{aa}
\addcontentsline{toc}{section}{\refname}\bibliography{my_bib}

\noindent \clearpage \onecolumn 
\appendix \section {Simplified half-integer expressions of the Einasto profile lensing properties} \label{Appendix-A}

\noindent The eqs. ~(\ref{Kappa-Meijer})-(\ref{flexion02-Meijer})
written in terms of Meijer $G$ function can be reduced to expressions
in terms of the generalized hypergeometric function \citep{9780124599017,1970hmfw.book.....A} 

\begin{equation}
_{p}F_{q}\left({a_{1},\ldots,a_{p};\, b_{1},\ldots,b_{q}};\, z\right)=\frac{\prod_{k=1}^{q}\Gamma(b_{k})}{\prod_{k=1}^{p}\Gamma(a_{k})}\,\frac{1}{2\pi i}\int_{{\cal \mathcal{L}}}\frac{\Gamma(s)\prod_{k=1}^{p}\Gamma(a_{k}-s)}{\prod_{k=1}^{q}\Gamma(b_{k}-s)}\,\left(-z\right)^{-s}\,{\text{d}}s,\label{eq:defpFq}
\end{equation}

\noindent for half-integer values of $n$, for which the poles are
simple, using the Slater's theorem \citep{marichev1983handbook}

\begin{equation}
G_{p,q}^{m,n}\left[\left.\begin{matrix}{\boldsymbol{a}}\\
{\boldsymbol{b}}
\end{matrix}\,\right|\, z\right]=\sum_{k=1}^{m}\frac{\prod_{j=1}^{m}\Gamma(b_{j}-b_{k})^{*}\prod_{j=1}^{n}\Gamma(1+b_{k}-a_{j})\, z^{b_{k}}}{\prod_{j=m+1}^{q}\Gamma(1+b_{k}-b_{j})\prod_{j=n+1}^{p}\Gamma(a_{j}-b_{k})}\,{}_{p}F_{q-1}\left[\left.\begin{matrix}{1+b_{k}-\boldsymbol{a}}\\
{\left(1+b_{k}-\boldsymbol{b}\right){}^{*}}
\end{matrix}\,\right|\, (-1)^{p-m-n} \; z\right],\label{eq:slater-theorem}
\end{equation}

\noindent where the asterisk in the gamma function indicates that
the term $k=j$, which corresponds to $\Gamma\left(0\right)$, must
be replaced by $1$; and in the hypergeometric function that the vector
$1+b_{k}-\boldsymbol{b}$ must be reduced in size from $q$ to $q-1$.

\noindent It is evident from eqs. ~(\ref{Kappa-Meijer}) and (\ref{eq:slater-theorem})
that the convergence may be written as 

\begin{equation}
\kappa\left(x\right)=\frac{\kappa_{\text{c}}}{2\,(2\pi)^{n-1}\,\sqrt{n}\,\Gamma\left(n\right)}\, x\,\left\{ \sum_{k=1}^{2n}\,\prod_{j=1}^{2n}\Gamma(b_{j}-b_{k})^{*}\,\left(\frac{x^{2}}{\left(2n\right)^{2n}}\right)^{b_{k}}\,{}_{0}F_{\,2n-1}\left[\left.\begin{matrix}-\\
{\left(1+b_{k}-\boldsymbol{b}\right){}^{*}}
\end{matrix}\,\right|\left(-1\right)^{-2n}\frac{x^{2}}{\left(2n\right)^{2n}}\right]\right\} ,\label{eq:convergence-hyperg}
\end{equation}

\noindent with $b_{k}$ the components of ${\boldsymbol{b}}$ given
by eq. (\ref{eq:vector-b}).

\noindent Inserting eq. (\ref{eq:convergence-hyperg}) into eqs. (\ref{eq:shear}),
(\ref{eq:1st-flex-pot-1}) and (\ref{eq:2nd-flex-pot-1}), we obtain
the expressions

\begin{multline}
\left|\gamma\left(x\right)\right|=\frac{-\kappa_{\text{c}}}{2\,(2\pi)^{n-1}\,\sqrt{n}\,\Gamma\left(n\right)}\, x\,\Biggl\{\sum_{k=1}^{2n}\,\prod_{j=1}^{2n}\Gamma(b_{j}-b_{k})^{*}\,\left(\frac{\frac{1}{2}+b_{k}}{\frac{3}{2}+b_{k}}\right)\,\left(\frac{x^{2}}{\left(2n\right)^{2n}}\right)^{b_{k}}\,\times\\
_{2}F_{\,2n+1}\left[\left.\begin{matrix}{\frac{3}{2}+b_{k},\frac{3}{2}+b_{k}}\\
{\frac{1}{2}+b_{k},\frac{5}{2}+b_{k},\left(1+b_{k}-\boldsymbol{b}\right){}^{*}}
\end{matrix}\,\right|\left(-1\right)^{-2n}\frac{x^{2}}{\left(2n\right)^{2n}}\right]\Biggr\},\label{eq:shear-hyperg}
\end{multline}

\begin{equation}
\left|\mathcal{F}\left(x\right)\right|=\frac{\mathcal{F}_{0}}{(2\pi)^{n-1}\,\sqrt{n\,\pi}}\,\left\{ \sum_{k=1}^{2n}\,\prod_{j=1}^{2n}\Gamma(b_{j}-b_{k})^{*}\,\left(\frac{1}{2}+b_{k}\,\right)\,\left(\frac{x^{2}}{\left(2n\right)^{2n}}\right)^{b_{k}}{}_{1}F_{\,2n}\left[\left.\begin{matrix}{\frac{3}{2}+b_{k}}\\
{\frac{1}{2}+b_{k},\left(1+b_{k}-\boldsymbol{b}\right){}^{*}}
\end{matrix}\,\right|\left(-1\right)^{-2n}\frac{x^{2}}{\left(2n\right)^{2n}}\right]\right\} ,\label{eq:flexion01-hyperg}
\end{equation}

\begin{multline}
\left|\mathcal{G}\left(x\right)\right|=\frac{\mathcal{F}_{0}}{(2\pi)^{n-1}\,\sqrt{n\,\pi}}\,\left\{ \sum_{k=1}^{2n}\,\prod_{j=1}^{2n}\Gamma(b_{j}-b_{k})^{*}\,\left(\frac{1}{2}+b_{k}\right)\,\left(\frac{x^{2}}{\left(2n\right)^{2n}}\right)^{b_{k}}\Biggl\{\,{}_{1}F_{\,2n}\left[\left.\begin{matrix}{\frac{3}{2}+b_{k}}\\
{\frac{1}{2}+b_{k},\left(1+b_{k}-\boldsymbol{b}\right){}^{*}}
\end{matrix}\,\right|\left(-1\right)^{-2n}\frac{x^{2}}{\left(2n\right)^{2n}}\right]\right.\\
\left.-\frac{2}{\frac{3}{2}+b_{k}}\,{}_{2}F_{\,2n+1}\left[\left.\begin{matrix}{\frac{3}{2}+b_{k},\frac{3}{2}+b_{k}}\\
{\frac{1}{2}+b_{k},\frac{5}{2}+b_{k},\left(1+b_{k}-\boldsymbol{b}\right){}^{*}}
\end{matrix}\,\right|\left(-1\right)^{-2n}\frac{x^{2}}{\left(2n\right)^{2n}}\right]\Biggr\}\right\} .\label{eq:flexion02-hyperg}
\end{multline}

\noindent Other important lensing quantities such as the cumulative
surface mass density $M\left(x\right)$, deflection angle $\alpha\left(x\right)$,
and deflection potential $\psi\left(x\right)$ also can be written
in terms of the generalized hypergeometric function

\begin{equation}
M\left(x\right)=\frac{\sqrt{n}\,\rho_{0}\, h^{3}}{2(2\pi)^{n-2}}\, x^{3}\,\left\{ \sum_{k=1}^{2n}\,\frac{\prod_{j=1}^{2n}\Gamma(b_{j}-b_{k})^{*}}{\left(\frac{3}{2}+b_{k}\right)}\,\left(\frac{x^{2}}{\left(2n\right)^{2n}}\right)^{b_{k}}\,_{1}F_{\,2n}\left[\left.\begin{matrix}{\frac{3}{2}+b_{k}}\\
{\frac{5}{2}+b_{k},\left(1+b_{k}-\boldsymbol{b}\right){}^{*}}
\end{matrix}\,\right|\left(-1\right)^{-2n}\frac{x^{2}}{\left(2n\right)^{2n}}\right]\right\} ,
\end{equation}

\begin{equation}
\alpha\left(x\right)=\frac{\kappa_{\text{c}}}{2\,(2\pi)^{n-1}\,\sqrt{n}\,\Gamma\left(n\right)}\, x^{2}\,\left\{ \sum_{k=1}^{2n}\,\frac{\prod_{j=1}^{2n}\Gamma(b_{j}-b_{k})^{*}}{\left(\frac{3}{2}+b_{k}\right)}\,\left(\frac{x^{2}}{\left(2n\right)^{2n}}\right)^{b_{k}}\,{}_{1}F_{\,2n}\left[\left.\begin{matrix}{\frac{3}{2}+b_{k}}\\
{\frac{5}{2}+b_{k},\left(1+b_{k}-\boldsymbol{b}\right){}^{*}}
\end{matrix}\,\right|\left(-1\right)^{-2n}\frac{x^{2}}{\left(2n\right)^{2n}}\right]\right\} ,
\end{equation}

\begin{multline}
\psi\left(x\right)=\frac{\kappa_{\text{c}}}{4\,(2\pi)^{n-1}\,\sqrt{n}\,\Gamma\left(n\right)}\, x^{3}\,\Biggl\{\sum_{k=1}^{2n}\,\frac{\prod_{j=1}^{2n}\Gamma(b_{j}-b_{k})^{*}}{\left(\frac{3}{2}+b_{k}\right)^{2}}\,\left(\frac{x^{2}}{\left(2n\right)^{2n}}\right)^{b_{k}}\,\times\\
_{2}F_{\,2n+1}\left[\left.\begin{matrix}{\frac{3}{2}+b_{k},\frac{3}{2}+b_{k}}\\
{\frac{5}{2}+b_{k},\frac{5}{2}+b_{k},\left(1+b_{k}-\boldsymbol{b}\right){}^{*}}
\end{matrix}\,\right|\left(-1\right)^{-2n}\frac{x^{2}}{\left(2n\right)^{2n}}\right]\Biggr\}.\label{eq:deflection-hyperg}
\end{multline}

\noindent The eqs. (\ref{eq:convergence-hyperg})-(\ref{eq:deflection-hyperg})
are ready for being use in numerical calculations because there are
already several available numerical implementations of the generalized
hypergeometric function. Some software that includes an implementation
of this function are the proprietary \texttt{{Maple}}, \texttt{{Mathematica}},
and \texttt{{Matlab}}, and freely available \texttt{{Sage}}, \texttt{{mpmath}}
library, and the package \texttt{{hypergeo} }\citep{Haking2006}
of the R language.

\noindent \section{Series expansions of the shear and first and second flexions} 
\label{Appendix-B}

\noindent The expressions for the shear $\gamma\left(x\right)$ and
first $\mathcal{F}(x)$ and second $\mathcal{G}(x)$ flexions can
be written as series expansions to study its asymptotic behaviour
near zero. We apply the residue theorem to the contour integral in
eq. (\ref{eq:defH}) and obtain the explicit power or power-logarithmic
series expansions depending on the multiplicity of the poles of the
gamma functions $\Gamma(b_{j}+B_{j}s)$. Examples of specific applications
of the residue theorem can be found in \citet{2011A&A...534A..69B},
for the deprojected S\'ersic profile, and in \citetalias{2012arXiv1202.5242R}
for the projected Einasto profile. The general theorem for the Fox
$H$ function can be found in \citet{KilbasSaigo99}. We encounter
two cases:

\noindent Case $1$: if $n$ is either non-rational or rational number
$p/q$ with an even denominator (and $p,q$ are coprime), all poles
are simple, so that the expansion takes the form of a power series

\begin{equation}
\left|\gamma(x)\right|=\frac{\kappa_{\text{c}}\,\sqrt{\pi}}{\Gamma\left(n\right)}\,\left\{ \,\sum_{k=1}^{\infty}\left[\left(\tfrac{1}{2}+\tfrac{k}{2n}\right)\,\frac{\Gamma\left(-\tfrac{3}{2}-\tfrac{k}{2n}\right)}{\Gamma\left(-\tfrac{k}{2n}\right)}\,\frac{(-1)^{k}}{k!}\,\frac{x^{k/n+1}}{2n}\right]\ -\sum_{k=0}^{\infty}\frac{\Gamma(n-2nk)}{\Gamma\left(\tfrac{1}{2}-k\right)}\,\frac{(-1)^{k}}{\left(k+1\right)\left(k-1\right)!}\, x^{2k}\,\right\} ,\label{eq:shear-gen-series}
\end{equation}

\begin{equation}
\left|\mathcal{F}(x)\right|=\mathcal{F}_{0}\,\left[\,-\sum_{k=1}^{\infty}\frac{\Gamma\left(\tfrac{1}{2}-\tfrac{k}{2n}\right)}{\Gamma\left(-\tfrac{k}{2n}\right)}\,\frac{(-1)^{k}}{k!}\,\frac{x^{k/n}}{n}\,+\,2\,\sum_{k=0}^{\infty}\frac{\Gamma(n-2nk)}{\Gamma\left(\tfrac{1}{2}-k\right)}\,\frac{(-1)^{k}}{\left(k-1\right)!}\, x^{2k-1}\,\right],\label{flexion01-gen-series-1}
\end{equation}

\noindent and

\begin{equation}
\left|\mathcal{G}(x)\right|=\mathcal{F}_{0}\,\left[\,\sum_{k=1}^{\infty}\frac{\Gamma\left(-\tfrac{1}{2}-\tfrac{k}{2n}\right)}{\Gamma\left(-\tfrac{k}{2n}\right)}\,\frac{k^{2}-n^{2}}{3n+k}\,\frac{(-1)^{k}}{k!}\,\frac{x^{k/n}}{2n^{2}}\,+\,2\,\sum_{k=0}^{\infty}\frac{\Gamma(n-2nk)}{\Gamma\left(\tfrac{1}{2}-k\right)}\,\frac{k-1}{k+1}\,\frac{(-1)^{k}}{\left(k-1\right)!}\, x^{2k-1}\,\right].\label{flexion02-gen-series}
\end{equation}

\noindent Case $2$: if $n$ is either an integer or rational number
$p/q$ with an odd denominator, some poles are of second order, then
the expansion takes the form of a logarithmic-power series

\begin{multline}
\left|\gamma(x)\right|=\frac{\kappa_{\text{c}}\,\sqrt{\pi}}{\Gamma\left(n\right)}\,\left\{ \sum_{\substack{k=1\\
k\,{\text{mod}}\, p\ne0
}
}^{\infty}\biggl[\left(\tfrac{1}{2}+\tfrac{k}{2n}\right)\,\frac{\Gamma\left(-\tfrac{3}{2}-\tfrac{k}{2n}\right)}{\Gamma\left(-\tfrac{k}{2n}\right)}\,\frac{(-1)^{k}}{k!}\,\frac{x^{k/n+1}}{2n}\biggr]\ -\ \hspace{-3ex}\sum_{\substack{k=1\\
(k+k_{0})\,{\text{mod}}\, q\ne0
}
}^{\infty}\hspace{-3ex}\frac{\Gamma(n-2nk)}{\Gamma\left(\tfrac{1}{2}-k\right)}\,\frac{(-1)^{k}}{\left(k+1\right)(k-1)!}\, x^{2k}\,\right\} \\
-\ \frac{\kappa_{\text{c}}}{n\,\Gamma\left(n\right)}\hspace{-3ex}\sum_{\substack{k=1\\
(k+k_{0})\,{\text{mod}}\, q=0
}
}^{\infty}\hspace{-3ex}\frac{(-1)^{p}\,(2k)!}{(2nk-n)!\,\left(k+1\right)!\,(k-1)!}\left(\frac{x}{2}\right)^{2k}\,\left[-\ln\left(\frac{x}{2}\right)-\frac{1}{2k}-\frac{1}{2k\left(k+1\right)}+\psi(k+1)+n\,\psi(2nk-n)-\psi(2k-1)\right],\label{shear-rat-series}
\end{multline}

\begin{multline}
\left|\mathcal{F}(x)\right|=\mathcal{F}_{0}\,\left[\ \ -\hspace{-1.5ex}\sum_{\substack{k=1\\
k\,{\text{mod}}\, p\ne0
}
}^{\infty}\hspace{-1.5ex}\frac{\Gamma\left(\tfrac{1}{2}-\tfrac{k}{2n}\right)}{\Gamma\left(-\tfrac{k}{2n}\right)}\,\frac{(-1)^{k}}{k!}\,\frac{x^{k/n}}{n}\ +\ 2\ \hspace{-3ex}\sum_{\substack{k=1\\
(k+k_{0})\,{\text{mod}}\, q\ne0
}
}^{\infty}\hspace{-3ex}\frac{\Gamma(n-2nk)}{\Gamma\left(\tfrac{1}{2}-k\right)}\,\frac{(-1)^{k}}{\left(k-1\right)!}\, x^{2k-1}\ \right]\\
+\ \frac{\mathcal{F}_{0}}{n\,\sqrt{\pi}}\hspace{-3ex}\sum_{\substack{k=1\\
(k+k_{0})\,{\text{mod}}\, q=0
}
}^{\infty}\hspace{-3ex}\frac{(-1)^{p}\,(2k)!}{(2nk-n)!\, k!\,(k-1)!}\left(\frac{x}{2}\right)^{2k-1}\,\left[-\ln\left(\frac{x}{2}\right)-\frac{1}{k}+\psi(k+1)+n\,\psi(2nk-n)-\psi(2k-1)\right],\label{flexion01-rat-series}
\end{multline}

\noindent and

\begin{multline}
\left|\mathcal{G}(x)\right|=\mathcal{F}_{0}\,\left\{ \sum_{\substack{k=1\\
k\,{\text{mod}}\, p\ne0
}
}^{\infty}\,\left[\frac{\Gamma\left(-\tfrac{1}{2}-\tfrac{k}{2n}\right)}{\Gamma\left(-\tfrac{k}{2n}\right)}\,\frac{k^{2}-n^{2}}{3n+k}\,\frac{(-1)^{k}}{k!}\,\frac{x^{k/n}}{2n^{2}}\right]\ +\ 2\ \hspace{-3ex}\sum_{\substack{k=1\\
(k+k_{0})\,{\text{mod}}\, q\ne0
}
}^{\infty}\hspace{-3ex}\frac{\Gamma(n-2nk)}{\Gamma\left(\tfrac{1}{2}-k\right)}\,\frac{k-1}{k+1}\,\frac{(-1)^{k}}{(k-1)!}\, x^{2k-1}\,\right\} \\
+\ \frac{\mathcal{F}_{0}}{n\,\sqrt{\pi}}\hspace{-3ex}\sum_{\substack{k=1\\
(k+k_{0})\,{\text{mod}}\, q=0
}
}^{\infty}\hspace{-3ex}\frac{(-1)^{p}\,(2k)!}{(2nk-n)!\,\left(k+1\right)!\,(k-2)!}\left(\frac{x}{2}\right)^{2k-1}\,\Bigl[-\ln\left(\frac{x}{2}\right)-\frac{1}{k-1}+\frac{1}{k\left(k+1\right)\left(k-1\right)}+\psi(k+1)+n\,\psi(2nk-n),\\
-\psi(2k-1)\Bigr],\label{flexion02-rat-series}
\end{multline}

\noindent with $\psi(k)$ the digamma function and $k_{0}=\tfrac{q-1}{2}$.
\end{document}